\documentclass[aps,prd,tightenlines,floats,twocolumn,nofootinbib,amsmath,amssymb,showpacs,superscriptaddress]{revtex4}
\usepackage{amsmath,amssymb,amsfonts}
\usepackage{graphicx}
\usepackage{enumerate} 
\usepackage{colordvi} 
\usepackage{bm}

\newcommand{\bfv}{\mbox{\boldmath$v$}}

\newcommand{\bfx}{\mbox{\boldmath$x$}}
\newcommand{\bfk}{\mbox{\boldmath$k$}}
\newcommand{\bfp}{\mbox{\boldmath$p$}}
\newcommand{\bfq}{\mbox{\boldmath$q$}}
\newcommand{\bfr}{\mbox{\boldmath$r$}}
\newcommand{\bfs}{\mbox{\boldmath$s$}}

\newcommand{\bfj}{\mbox{\boldmath$j$}}
\newcommand{\bfA}{\mbox{\boldmath$A$}}

\newcommand{\Pdd}{P_{\delta\delta}}
\newcommand{\Pdv}{P_{\delta \theta}}
\newcommand{\Pvv}{P_{\theta\theta}}
\newcommand{\DFoG}{D_{\rm FoG}}

\newcommand{\sigmav}{\sigma_{\rm v}}

\begin{document}
\title{Precision modeling of redshift-space distortions from multi-point propagator expansion} 

\author{Atsushi Taruya}
\affiliation{Research Center for the Early Universe, School of Science, 
The University of Tokyo, Bunkyo-ku, Tokyo 113-0033, Japan}
\affiliation{
Kavli Institute for the Physics and Mathematics of the Universe, Todai Institutes for Advanced Study, the University of Tokyo, Kashiwa, Chiba 277-8583, Japan (Kavli IPMU, WPI)}
\author{Takahiro Nishimichi}
\affiliation{
Kavli Institute for the Physics and Mathematics of the Universe, Todai Institutes for Advanced Study, the University of Tokyo, Kashiwa, Chiba 277-8583, Japan (Kavli IPMU, WPI)}
\author{Francis Bernardeau}
\affiliation{Institut de Physique Th\'eorique, CEA, F-91191
  Gif-sur-Yvette, France\\
  \ \ CNRS, URA 2306, F-91191, Gif-sur-Yvette, France}
\begin{abstract}
Using a full implementation of resummed perturbation theory (PT) from
a multi-point propagator expansion, we put forward new theoretical 
predictions for the two-point statistics of matter fluctuations in redshift space. 
The predictions consistently include PT corrections up to the two-loop 
order and are based on an improved prescription of the redshift-space distortions (RSD) that  properly 
takes into account their non-Gaussian impact 
from  a systematic low-$k$ expansion. 
In contrast to the previous studies that partly used standard PT calculations, the present treatment is able to provide 
a consistent prediction for both power spectra and correlation functions. These results are  
compared with $N$-body simulations with which a very good agreement 
is found up to the quadrupole moment. 
The theoretical predictions for the hexadecapole moment of
the power spectra are however found 
to significantly departs from the numerical results at low redshift.  We examine this issue  and 
found it to be likely related to an improper modeling of the RSD damping effects
on which this moment shows large dependence. 
\end{abstract}

\maketitle

\section{Introduction}
\label{sec:intro}

The large-scale structure of the Universe observed via galaxy redshift 
surveys appears distorted due to the peculiar velocity of galaxies, 
known as the redshift-space distortions (RSD) (e.g., 
\cite{Hamilton:1997zq,Peebles:1980}). The RSD
breaks the statistical isotropy, and 
the resultant galaxy clustering exhibits characteristic anisotropies 
along observer's line-of-sight direction 
by the two competitive effects, i.e., Kaiser and Finger-of-God effects 
\cite{Kaiser:1987qv,1972MNRAS.156P...1J,1983ApJ...267..465D,
Scoccimarro:2004tg}. While the latter effect mostly comes from the 
virialized random motion of the mass (or galaxy) in halos, 
the Kaiser effect apparently enhances the clustering amplitude along 
a line-of-sight direction, and the effect is dominated by 
the large-scale coherent motion. In particular, the strength of the 
Kaiser effect is simply described by the linear theory, and is characterized 
by the growth-rate parameter, $f$, defined as $f=d\ln D_+/d\ln a$, where 
the quantities $D_+$ and $a$ are the linear growth factor and scale factor 
of the Universe, respectively (e.g., \cite{1992ApJ385L5H,Cole:1993kh}). 
Thus, the Kaiser effect can be 
used as a useful tool to measure the growth of cosmological structure, 
and combining the distance measurements, the measurement of RSD 
offers a unique opportunity to test the gravity theory on cosmological scales 
(e.g., \cite{Linder:2007nu,Guzzo:2008ac,Yamamoto:2008gr,
Song:2008qt,Song:2010bk}).  
Note that the galaxy redshift surveys also provide a 
way to measure the cosmological distance, utilizing the so-called 
Alcock-Paczynski effect \cite{Alcock_Paczynski:1979}. With the 
baryon acoustic oscillations (BAOs) imprinted on the large-scale structure
as a robust standard ruler, we can thus measure
the angular diameter distance $D_A(z)$ and the Hubble parameter 
$H(z)$ of distant galaxies at redshift $z$ separately 
through the Alcock-Paczynski effect (e.g., \cite{Seo:2003pu,
Blake:2003rh,Glazebrook:2005mb,Shoji:2008xn,Padmanabhan:2008ag}).

Hence, the large-scale galaxy distribution offers a dual cosmological probe
to simultaneously constrain both the cosmic expansion and structure 
growth, from which we can directly test gravity, and even 
address the nature of late-time cosmic acceleration (see 
\cite{Reid:2009xm,Percival:2009xn,Kazin:2010nd,Yamamoto:2008gr,
Reid:2012sw} for recent status). This is 
the main reason why  ongoing and planned galaxy surveys 
aim at precisely measuring the RSD and BAOs through the 
clustering statistics of galaxy distribution. On the other hand, 
a pursuit of such a precision measurement poses several challenging issues 
in theory of large-scale structure. One important issue is 
the development of precision theoretical tools to compute 
the clustering statistics of the 
large-scale structure. While the RSD and BAO are measured
from the galaxy clustering at the scales close to the linear regime of 
the gravitational evolution, nonlinearity of both the gravity and the 
RSD is known to play a crucial role in precise estimate of the parameters 
$f$, $D_A$, and $H$ (e.g., \cite{Taruya:2010mx,Taruya:2011tz,
Nishimichi:2011jm}).

The aim of the present paper is to address such an 
issue and to investigate the extent to which we can accurately compute 
the redshift-space power spectrum and correlation function based 
on the perturbation theory (PT). In redshift space, a key point is 
that the applicable range of linear theory prediction is fairly 
narrower than that in real space, and the corrections coming from 
the nonlinear clustering and RSD need to be properly incorporated into 
theoretical predictions even on such large-scales as $k\lesssim0.1\,h$Mpc$^{-1}$. 
This is because, in terms of real-space quantities, 
the redshift-space power spectrum and/or correlation function 
cannot simply be expressed as the large-scale two-point statistics of the underlying fields
and are significantly  affected by the small scale 
physics\cite{Scoccimarro:2004tg}. 
Thus, for a reliable theoretical predictions with 
a wider applicable range, a sophisticated treatment of 
both the gravitational clustering and RSD is required.

In our previous studies, we have proposed an improved model of RSD relevant 
in the weakly nonlinear regime \cite{Taruya:2010mx,Nishimichi:2011jm} 
(see \cite{Matsubara:2007wj,Reid:2011ar,Carlson:2012bu,Seljak:2011tx,Vlah:2012ni} for other improved models or treatments). 
The model properly accounts of the non-Gaussian 
nature of the RSD based on the low-$k$ expansion. The resulting 
power spectrum expression differs from the one of the so-called streaming 
model frequently used in the literature (e.g., \cite{Peebles:1980,Hatton:1997xs,Scoccimarro:2004tg,Jennings:2010ne}) with the introduction
of additional corrections associated with 
nonlinear couplings between velocity and density fields \cite{Taruya:2010mx}. 
A detailed investigation revealed that these corrections can give an important contribution 
 to the acoustic structure of BAOs which gives rise to a slight increase in 
the amplitude of monopole and quadrupole spectra. While 
the model has been originally proposed for the matter power spectrum, 
with an improved PT of gravitational clustering as well as an 
appropriate parameterization of galaxy/halo bias, it has been 
shown to successfully describe not only the matter but also the 
halo power spectra in $N$-body simulations \cite{Nishimichi:2011jm}.

In this paper, the third of a series on this method,  
we present consistent PT calculations aiming at giving predictions in both 
Fourier and configuration spaces.  In all of our previous works, the 
PT predictions have been done in some heterogeneous ways. That is, 
while the improved PT scheme by Ref.~\cite{Taruya:2007xy,Taruya:2009ir} 
is applied to compute the power spectra for 
density and velocity fields,  
the standard PT treatment is partly 
used to compute the new corrections of the RSD, for which we have only 
given the predictions based on the leading-order PT calculation. 
It is generally known that the standard PT treatment produces an ill-behaved 
PT expansion that has poor convergence properties
(e.g., \cite{Crocce:2005xy,Carlson:2009it,Taruya:2009ir}). 
This is indeed the case 
when we compute the redshift-space power spectrum. 
Because of the bad UV behavior, 
the correction terms computed with standard PT eventually become dominant 
at small scales. Then, a naive computation of the correlation function 
through the direct integration of the power spectrum 
ceases to give a physically reasonable answer. This is one of  
the main reasons why the previous works have focused on the 
redshift-space power spectrum, not the correlation function.

Here, in order to remedy the bad UV behaviors, we will apply 
the specific resummed PT scheme referred to as 
the multi-point propagator expansion or $\Gamma$ expansion 
\cite{Bernardeau:2008fa}. 
The advantage of this scheme is that the non-perturbative properties, 
which can be obtained in standard PT by summing up infinite series of PT 
expansions, are whole encapsulated in the multi-point propagators, with which  
any statistical quantities including the power spectrum, 
bispectrum and trispectrum can be built up. In particular, 
the asymptotic behaviors of the multi-point propagators are analytically 
known \cite{Bernardeau:2008fa,Bernardeau:2011vy}, 
and with a help of these analytic properties, a novel regularized treatment
has been proposed \cite{Bernardeau:2011dp}, 
allowing us to consistently interpolate between 
standard PT results at low-$k$ and the expected resummed behavior at 
high-$k$. In this paper, making full use of the regularized 
$\Gamma$ expansion, 
we are now able to give a consistent calculation for both the power spectrum 
and correlation function in redshift space. 
We will present the results including the PT corrections up to the 
next-to-next-to-leading order, i.e. two-loop order, and compare those 
with $N$-body simulations in detail. With the calculations at the two-loop 
order, we will also discuss the potential impact of the 
higher-order non-Gaussian corrections coming from the RSD. 
While the model of RSD has been derived based on the low-$k$ 
expansion, one of the corrections, comparable to the two-loop order in 
PT expansion, has been ignored in previous studies without any justification. 
Here, we will explicitly quantify the size of this correction, and with a 
help of the $N$-body simulations, we will explore the potential 
systematics of our model predictions.

The paper is organized as follows. In Sec.~\ref{sec:TNS_model}, 
we begin by briefly reviewing the model of RSD. 
Sec.~\ref{sec: pk_bk_from_Gamma_expansion} describes the multi-point 
propagator expansion which we will apply to the predictions of 
redshift-space power spectrum and correlation function. Based on the 
regularized scheme, the basic formalism to compute the propagators 
and the statistical quantities built up with those is presented, 
together with technical detail of the implementation. Then, 
Sec.~\ref{sec:results} presents the main results. The PT predictions up 
to the two-loop order are compared with $N$-body simulations. 
In Sec.~\ref{sec:discussion}, the impact of higher-order corrections of 
RSD is discussed, and the correction that has been so far neglected is 
explicitly computed. With a help of $N$-body simulations, a potential 
systematics in the model prediction is also investigated. 
Finally, Sec.~\ref{sec:conclusion} is devoted to the summary and 
conclusions.

\section{The model of redshift-space distortions}
\label{sec:TNS_model}

Here, we briefly review the model prescription of redshift-space distortions 
(RSD) proposed by Ref.~\cite{Taruya:2010mx}, 
and discuss several remarks on the 
perturbation theory (PT) treatment 
in computing redshift-space power spectrum or correlation function.

We begin by writing the exact expression for redshift-space power spectrum. 
Let us denote the density and velocity fields by $\delta$ and $\bfv$. 
Owing to the distant-observer approximation, which is usually valid for 
the observation of distant galaxies of our interest, one can write 
(e.g., \cite{Scoccimarro:2004tg,Bernardeau:2001qr,Taruya:2010mx})
\begin{align}
&P^{\rm(S)}(\bfk)=\int d^3\bfx\,e^{i\bfk\cdot\bfx}
\bigl\langle e^{-ik\mu\,f \Delta u_z}
\nonumber\\
&\qquad\quad\times
\left\{\delta(\bfr)+f\nabla_zu_z(\bfr)\right\}
\left\{\delta(\bfr')+f\nabla_zu_z(\bfr')\right\}\bigr\rangle,
\label{eq:Pkred_exact}
\end{align}
where $\bfx=\bfr-\bfr'$ denotes the separation in real space 
and $\langle\cdots\rangle$ indicates an ensemble average. 
In the above expression, the $z$-axis is taken as 
observer's line-of-sight direction, and we define the directional cosine 
$\mu$ by $\mu=k_z/k$. Further, 
we defined $u_z(\bfr)=-v_z(\bfr)/(aHf)$, and $\Delta u_z=u_z(\bfr)-u_z(\bfr')$
for the line-of-sight component of the velocity field. 
Here, the function $f$ is the 
logarithmic derivative of the linear growth function $D(z)$ given by 
$f=d\ln D(z)/d\ln a$. Note that the above expression has been derived 
without invoking the dynamical information 
for velocity and density fields, i.e., the Euler equation and/or continuity 
equation.

Clearly, 
the redshift-space power spectrum $P^{\rm(S)}$, 
given as the function of wavenumber $k$ and the directional cosine $\mu$,   
cannot be derived from the mere two-point statistics of the underlying fields. 
If we expand the exponent in the 
bracket, it can be described by the infinite series of multi-point 
spectra of density and velocity fields. 
In order to derive an analytically tractable expression applicable to 
the large-scale cosmological observations, 
we rewrite the expression (\ref{eq:Pkred_exact}) with 
\begin{equation}
P^{\rm(S)}(k,\mu)=\int d^3\bfx\,e^{i\,\bfk\cdot\bfx}
\bigl\langle e^{j_1A_1}A_2A_3\bigr\rangle, 
\label{eq:Pks_exact}
\end{equation}
where the quantities $j_1$, $A_i\,(i=1,2,3)$ are respectively given by 
\begin{align}
&j_1= -i\,k\mu f,\nonumber\\
&A_1=u_z(\bfr)-u_z(\bfr'),\nonumber\\
&A_2=\delta(\bfr)+\,f\nabla_zu_z(\bfr),\nonumber\\
&A_3=\delta(\bfr')+\,f\nabla_zu_z(\bfr').
\label{eq:j_A}
\end{align}
We shall rewrite the ensemble average 
$\langle e^{j_1A_1}A_2A_3\rangle$ in terms of cumulants. 
To do this, we use the relation between the cumulant and 
moment generating functions. For the stochastic vector field 
$\bfA=\{A_1,A_2,A_3\}$, we have 
(e.g., \cite{Scoccimarro:2004tg, Matsubara:2007wj}): 
\begin{equation}
\langle e^{\bfj\cdot\bfA}\rangle=
\exp \left\{\langle e^{\bfj\cdot\bfA}\rangle_c\right\}
\end{equation}
with $\bfj$ being an arbitrary constant vector, $\bfj=\{j_1,j_2,j_3\}$. 
Taking the derivative twice with 
respect to $j_2$ and $j_3$, we then set $j_2=j_3=0$. We obtain 
\cite{Scoccimarro:2004tg}
\begin{align}
&\langle e^{j_1A_1}A_2A_3\rangle=
\exp \left\{\langle e^{j_1A_1}\rangle_c\right\}
\nonumber\\
&\qquad\times
\left[\langle e^{j_1A_1}A_2A_3 \rangle_c+ 
\langle e^{j_1A_1}A_2\rangle_c \langle e^{j_1A_1}A_3 \rangle_c \right].
\end{align}
Substituting this into Eq.(\ref{eq:Pks_exact}), we arrive at
\begin{align}
&P^{\rm(S)}(k,\mu)=\int d^3\bfx \,\,e^{i\bfk\cdot\bfx}\,\,
\exp \left\{\langle e^{j_1A_1}\rangle_c\right\}
\nonumber\\
&\quad\quad
\times\left[\langle e^{j_1A_1}A_2A_3 \rangle_c+ 
\langle e^{j_1A_1}A_2\rangle_c \langle e^{j_1A_1}A_3 \rangle_c \right].
\label{eq:Pkred_exact2}
\end{align}
This expression clearly reveals the coupling between density and velocity 
fields associated with the Kaiser and Finger-of-God effects. 
In addition to the prefactor $\exp \left\{\langle e^{j_1A_1}\rangle_c\right\}$, 
the ensemble averages over the quantities $A_2$ and $A_3$ 
responsible for the Kaiser effect 
all include the exponential factor $e^{j_1A_1}$, which can produce 
a non-negligible correlation between density and velocity.

Based on Eq.~(\ref{eq:Pkred_exact2}), 
the proposition raised by Ref.~\cite{Taruya:2010mx} is the following:

\begin{itemize}
\item The exponential prefactor, 
$\exp \left\{\langle e^{j_1A_1}\rangle_c\right\}$, which is responsible for 
the suppression of power spectrum 
due to the virialized random and coherent motion, 
turns out to mainly affect the broadband shape of the power spectrum 
(Sec.III-B of \cite{Taruya:2010mx}). Nevertheless, the effect of this 
is known to be significant on BAO scales, and seems 
difficult to treat perturbatively. Thus, 
we replace the exponential prefactor with general functional form, 
$D_{\rm FoG}(k\mu\,f\,\sigmav)$ with $\sigmav$ being a constant, 
just ignoring the spatial correlation of $A_1$.  

\item In the bracket, the exponential factor $e^{j_1A_1}$ is very likely to 
affect the structure of BAO in the power spectrum, since the bracket 
includes the term leading to the Kaiser effect in linear regime. 
At the scale of our interest, the contribution coming from the factor 
$e^{j_1A_1}$ should be basically small, and the perturbative expansion may 
work well. Regarding the quantity $j_1$ as a small expansion parameter, we 
perturbatively expand the terms in the bracket of the integrand. Up to 
the second order in $j_1$, we have
\begin{align}
&\langle e^{j_1A_1}A_2A_3 \rangle_c+
\langle e^{j_1A_1}A_2\rangle_c \langle e^{j_1A_1}A_3 \rangle_c 
\nonumber\\
&\,\,\,\simeq\langle A_2A_3\rangle_c + j_1\langle A_1A_2A_3\rangle_c
\nonumber\\
&\quad
+j_1^2\Bigl\{ \frac{1}{2}\,\langle A_1^2A_2A_3\rangle_c 
+\langle A_1A_2\rangle_c\langle A_1A_3\rangle_c
\Bigr\}+\mathcal{O}(j_1^3).
\label{eq:low-k_expansion}
\end{align}
\end{itemize}

Then, from Eq.~(\ref{eq:Pkred_exact2}), the proposed model of RSD is 
expressed as
\begin{align}
&P^{\rm(S)}(k,\mu)=\DFoG[k\mu\,f\sigmav]\,
\nonumber\\
&~\times\Bigl\{P_{\rm Kaiser}(k,\mu)+A(k,\mu)+B(k,\mu)+D(k,\mu) 
\Bigr\}. 
\label{eq:TNS_model}
\end{align}
Specific form of the damping function $D_{\rm FoG}$ will be given later 
[see Eq.(\ref{eq:Damping_func})]. 
Here, the power spectrum $P_{\rm Kaiser}$ is originated from the term 
$\langle A_2A_3\rangle_c$, and it corresponds to
the nonlinear generalization of 
the Kaiser formula frequently used in the literature (e.g., 
\cite{Scoccimarro:2004tg,Percival:2008sh}): 
\begin{align}
P_{\rm Kaiser}(k,\mu)=\Pdd(k)
+2\,f\mu^2\,\Pdv(k)
+f^2\mu^4\,\Pvv(k).
\label{eq:Kaiser}
\end{align}
The functions $\Pdd$, $\Pdv$, and $\Pvv$ are the auto-power spectra of 
density and velocity divergence, and their cross spectrum. The 
velocity divergence is defined by $\theta=-\nabla\bfv/(f\,aH)$. 
On the other hand, the terms $A$, $B$ and $D$ induce the 
corrections arising from the low-$k$ expansion, and these are 
respectively defined by
\begin{eqnarray}
  \label{eq:correction}
  A(k,\mu)&=& j_1\,\int d^3\bfx \,\,e^{i\bfk\cdot\bfx}\,\,
\langle A_1A_2A_3\rangle_c,
\label{eq:A}\\
  B(k,\mu)&=& j_1^2\,\int d^3\bfx \,\,e^{i\bfk\cdot\bfx}\,\,
\langle A_1A_2\rangle_c\,\langle A_1A_3\rangle_c.
\label{eq:B}\\
  D(k,\mu)&=& \frac{j_1^2}{2}\,\int d^3\bfx \,\,e^{i\bfk\cdot\bfx}\,\,
\langle A_1^2A_2A_3\rangle_c.
\label{eq:D}
\end{eqnarray}
The term $D$ turns out to be of higher order if we employ the perturbation 
theory calculation, and in Ref.~\cite{Taruya:2010mx}, 
it has been neglected. While we basically follow 
their treatment here ignoring the $D$ term, as already discussed in 
Ref.~\cite{Taruya:2010mx}, this is a priori no longer consistent at two-loop 
order. The validity of such an heterogeneous 
treatment, and specifically the impact of $D$ term on the predictions of 
redshift-space power spectrum will be later discussed in detail 
(see Sec.~\ref{sec:discussion}).

In computing each term of the expression (\ref{eq:TNS_model}) with the PT treatment of large-scale structure, we invoke a single-stream approximation, 
in which the dynamics of 
large-scale structure is described by the density $\delta$ and 
velocity divergence $\theta$. 
The expression in Eq.~(\ref{eq:Kaiser}) is also the outcome of 
the single-stream approximation, and it is usually
valid as long as we are interested in the linear and weakly nonlinear 
regimes of the gravitational clustering. Then, the expressions for the 
terms $A$ and $B$ can be recast as
\begin{align}
&A(k,\mu)= (k\mu\,f)\,\int \frac{d^3\bfp}{(2\pi)^3} \,\,\frac{p_z}{p^2}
\nonumber\\
&\qquad\quad\times
\left\{B_\sigma(\bfp,\bfk-\bfp,-\bfk)-B_\sigma(\bfp,\bfk,-\bfk-\bfp)\right\},
\label{eq:A_term}
\\
&B(k,\mu)= (k\mu\,f)^2\int \frac{d^3\bfp}{(2\pi)^3} F(\bfp)F(\bfk-\bfp)\,\,;
\label{eq:B_term}
\\
&\qquad\quad F(\bfp)=\frac{p_z}{p^2}
\left\{ \Pdv(p)+\,f\frac{p_z^2}{p^2}\,\Pvv(p)\,\right\},
\nonumber
\end{align}
where the function $B_\sigma$ is 
the cross bispectra defined by 
\begin{align}
&\left\langle \theta(\bfk_1)
\left\{\delta(\bfk_2)+f\,\frac{k_{2z}^2}{k_2^2}\theta(\bfk_2)\right\}
\left\{\delta(\bfk_3)+f\,\frac{k_{3z}^2}{k_3^2}\theta(\bfk_3)\right\}
\right\rangle
\nonumber\\
&\quad\qquad
=(2\pi)^3\delta_D(\bfk_1+\bfk_2+\bfk_3)\,B_\sigma(\bfk_1,\bfk_2,\bfk_3).
\label{eq:def_B_sigma}
\end{align}

Note that  while we employed 
the low-$k$ expansion, we do not assume that the terms $A_i$ themselves are 
entirely small. In this sense, the expression (\ref{eq:TNS_model}) with 
the corrections (\ref{eq:A_term}) and (\ref{eq:B_term}) 
still holds some non-perturbative properties. 
A more detailed study revealed 
that the $A$ and $B$ terms basically give the positive 
contributions, and moderately but notably affect 
the shape and structure of BAOs. In particular, 
as revealed by Ref.~\cite{Nishimichi:2011jm}, the $A$ term exhibits a 
strong dependence on the halo/galaxy biasing, 
leading to a large-scale enhancement in 
amplitude relative to the real-space clustering. 
The effect is especially 
prominent for massive halos or highly biased objects, and in the presence 
of $A$ and $B$ terms, the model (\ref{eq:TNS_model}) indeed reproduces the   
halo redshift-space clustering quite well.

In this paper, based on the $\Gamma$-expansion, 
we will make a fully consistent calculation of the redshift-space 
power spectrum [Eq.~(\ref{eq:TNS_model})], including the 
PT corrections up to the two-loop order. With the regularization proposed 
in Ref.~\cite{Bernardeau:2011dp}, the power spectrum can be computed 
with a well-behaved UV behavior, which enables us to give a quantitative
predictions for the correlation function.

Finally, we briefly mention other improved models and treatments 
proposed recently, and note their qualitative differences. 
Ref.~\cite{Reid:2011ar} proposed the PT model based 
on the streaming model, allowing the scale-dependent velocity dispersion. 
Including the non-Gaussian corrections computed with standard PT,  
the model successfully describes the anisotropic correlation functions. 
Accurate prescriptions for the anisotropic correlation functions have been 
also given in Refs.~\cite{Matsubara:2007wj,Carlson:2012bu},  but 
these are constructed based on the Lagrangian PT. On the other hand, 
Ref.~\cite{Vlah:2012ni} presented an alternative PT prescription for 
redshift-space power spectrum, based on the moment-based expansion 
proposed by Ref.~\cite{Seljak:2011tx}.  
In this treatment, the higher-order corrections of RSD 
are all measurable quantities in the $N$-body simulation 
\cite{Okumura:2011pb,Okumura:2012xh}, and 
Ref.~\cite{Vlah:2012ni} compared their PT results with simulations 
term by term. 
Incorporating the effects of the small-scale velocity dispersions, 
the model has been shown to accurately describe the power spectrum in 
weakly nonlinear regime. Our RSD model 
may be regarded as a semi-PT model in the sense that a part of 
the terms (i.e., damping function) is not perturbatively treated, 
introducing a free parameter. Nevertheless, 
with the $\Gamma$ expansion, the model for the first time 
gives a consistent and accurate prediction 
for both the power spectra and correlation functions.

\section{The $\Gamma$ expansion and computation of redshift-space power 
spectrum}
\label{sec: pk_bk_from_Gamma_expansion}

The expression of redshift-space power spectrum in previous
section involves not only the real-space power spectra but also the 
higher-order corrections such as bispectrum. Although our main focus
is the weakly non-linear regime of gravitational clustering, the 
standard PT is known to produce ill-behaved higher-order corrections, 
that lack good convergence properties. 
Therefore, practical calculations of redshift-space power spectrum are better made 
with a resummed PT scheme, improving the convergence of PT expansion 
on small scales, so that correlation functions can be safely computed.
In this paper, we consider  the $\Gamma$ expansion, adopting the prescription for the regularization 
PT calculation by Ref.~\cite{Bernardeau:2011dp}.

\subsection{Regularized $\Gamma$ expansion}
\label{subsec:Gamma_expansion}

As seen in previous section, the density $\delta$ and velocity divergence 
$\theta$ play an important role to describe the redshift-space 
power spectrum. Let us introduce the two-component multiplet: 
\begin{align}
\Psi_a(\bfk)=\left(\delta(\bfk),\,\,\frac{\theta(\bfk)}{f}\right), 
\label{eq:doublet}
\end{align}
where the subscript $a=1,\,2$ selects the density and the velocity 
components. For our interest of the weakly nonlinear scales, where the 
single-stream approximation gives a very accurate prescription, 
the evolution of $\Psi_a$ is governed by the dynamics of 
the self-gravitating pressureless and irrotational fluid flow 
\cite{Bernardeau:2001qr}. 
To perturbatively solve the equation for fluid dynamics, 
a naive treatment with the standard PT is to just expand the fields $\Psi_a$ 
in terms of the 
initial fields. For the late-time epoch at which the growing-mode contribution 
is dominant, we then formally obtain the following expression: 
\begin{align}
&\Psi_a(\bfk)=\sum_{n=1}^\infty e^{n\,\eta} 
\int\frac{d^3\bfk_1\cdots d^3\bfk_n}{(2\pi)^{3(n-1)}}
\delta_{\rm D}(\bfk-\bfk_{1\cdots n})
\nonumber\\
&\quad\qquad\qquad \times F_a^{(n)}(\bfk_1,\cdots,\bfk_n)
\,\delta_0(\bfk_1)\cdots\delta_0(\bfk_n), 
\end{align}
where $\delta_0$ is the initial density field, and $\eta=\ln D(t)$ with
the quantity $D$ being the linear growth factor. The kernel $F_a^{(n)}$ 
is the symmetric function, and sometimes written as $F_a^{(n)}=(F_n,\,G_n)$, 
whose explicit expressions can be recursively obtained \cite{Bernardeau:2001qr}.

As we mentioned in Sec.~\ref{sec:intro}, however, the standard PT is known
to produce ill-behaved high-$k$ behavior. This prevents us from 
obtaining a convergent result for the correlation function. 
Here, as alternative to the standard PT framework, 
we consider the $\Gamma$ expansion, which is one of the non-perturbative PT  
frameworks. In this scheme, the multipoint propagator constitute 
the building blocks, and all the statistical quantities can be expressed in terms of these propagators. Denoting the $(p+1)$-point propagator by 
$\Gamma^{(p)}$, we define
\begin{align}
&\frac{1}{p!}\left\langle
\frac{\delta^p\Psi_a(\bfk,\eta)}{\delta\delta_0(\bfk_1)
\cdots \delta\delta_0(\bfk_p)}\right\rangle =\delta_{\rm D}
(\bfk-\bfk_{1\cdots p})
\nonumber\\
&\qquad\qquad\qquad\times\frac{1}{(2\pi)^{3(p-1)}} \,
\Gamma_a^{(p)}(\bfk_1,\cdots,\bfk_p;\eta).  
\end{align}
With these objects, the power spectra are shown to be expressed as 
\cite{Bernardeau:2008fa},
\begin{align}
&P_{ab}(|\bfk|;\eta) = 
\sum_{t=1}^{\infty}
t!\int\frac{d^3\bfq_1\cdots d^3\bfq_t}
{(2\pi)^{3(t-1)}}\,\,\delta_{\rm D}(\bfk-\bfq_{1\cdots t})\,
\nonumber\\
&\times\Gamma_a^{(t)}(\bfq_1,\cdots,\bfq_t;\eta)
\Gamma_b^{(t)}(\bfq_1,\cdots,\bfq_t;\eta)\,P_0(q_1)\cdots P_0(q_t). 
\label{eq:G_expansion_Pk}
\end{align}
Further, the bispectrum is expressed as
\begin{widetext}
\begin{align}
&B_{abc}(\bfk_1, \bfk_2, \bfk_3;\eta) = 
\sum_{r,s,t}\left(
\begin{array}{c}
r+s
\\
r
\end{array}
\right)\left(
\begin{array}{c}
s+t
\\
s
\end{array}
\right)\left(
\begin{array}{c}
t+r
\\
t
\end{array}
\right)
r! \,s!\,t!
\nonumber\\
&\times\,\int\frac{d^3\bfp_1\cdots d^3\bfp_r}{(2\pi)^{3(r-1)}}
\frac{d^3\bfq_1\cdots d^3\bfq_s}{(2\pi)^{3(s-1)}}
\frac{d^3\bfr_1\cdots d^3\bfr_t}{(2\pi)^{3(t-1)}}
\,\,\delta_{\rm D}(\bfk_1-\bfp_{1\cdots r}-\bfq_{1\cdots s})\,
\delta_{\rm D}(\bfk_2+\bfq_{1\cdots s}-\bfr_{1\cdots t})\,
\delta_{\rm D}(\bfk_3+\bfr_{1\cdots t}-\bfp_{1\cdots r})\,
\nonumber\\
&\times \Gamma_a^{(r+s)}(\bfp_1,\cdots,\bfp_r,\bfq_1,\cdots,\bfq_s;\eta)
\Gamma_b^{(s+t)}(-\bfq_1,\cdots,\bfq_s,\bfr_1,\cdots,\bfr_t;\eta)\,
\Gamma_c^{(t+r)}(-\bfr_1,\cdots,-\bfr_t,-\bfp_1,\cdots,-\bfp_r;\eta)\,
\nonumber\\
&\times P_0(p_1)\cdots P_0(p_r)P_0(q_1)\cdots P_0(q_s)P_0(r_1)\cdots P_0(r_t).
\label{eq:G_expansion_Bk}
\end{align}
\end{widetext}

The multipoint propagators are the non-perturbative quantities that can 
be obtained by summing up a class of infinite series of the standard PT 
expansion. The important remark is that one can exploit the asymptotic 
properties of the propagator $\Gamma^{(p)}$ in both low- and high-$k$ regimes. 
To be precise, in the high-$k$ limit,
higher-order contributions can be systematically computed at all orders, and 
as a result of summing up all the contributions,  the multi-point propagators 
are shown to be exponentially suppressed~\cite{Bernardeau:2008fa,Bernardeau:2011vy},
\begin{align}
&\Gamma^{(p)}_a(\bfk_1,\cdots,\bfk_p;\eta)
\nonumber\\
&\qquad
\stackrel{k\to\infty}{\longrightarrow} 
\exp\left\{-\frac{k^2\sigmav^2e^{2\eta}}{2}\right\}
\Gamma^{(p)}_{a,{\rm tree}}(\bfk_1,\cdots,\bfk_p;\eta)
\label{eq:Gamma_high-k}
\end{align}
with $k=|\bfk_1+\cdots+\bfk_p|$. Here, the quantity 
$\Gamma^{(p)}_{a,{\rm tree}}$ is the lowest-order 
non-vanishing propagator obtained from the standard PT calculation, and 
$\sigmav$ is the one-dimensional root-mean-square of the displacement 
field defined by,
\begin{align}
\sigmav^2=\frac{1}{3}\int\frac{d^3\bfq}{(2\pi)^3}\,\frac{P_0(q)}{q^2}.
\label{eq:sigam_v}
\end{align}
On the other hand, at low-$k$, the propagators are expected to approach 
their standard PT expressions that can be written formally,
\begin{align}
&\Gamma_a^{(p)}(\bfk_1,\cdots,\bfk_p;\eta)
=\Gamma_{a,{\rm tree}}^{(p)}(\bfk_1,\cdots,\bfk_p;\eta)
\nonumber\\
&\qquad\qquad
+\sum_{n=1}^\infty\Gamma_{a,n\mbox{-}{\rm loop}}^{(p)}(\bfk_1,\cdots,\bfk_p;\eta).
\label{eq:Gamma_loop_expand}
\end{align}
For the dominant growing-mode contribution we are interested in, 
each correction term is expressed in terms of the standard PT kernels as,
\begin{align}
\Gamma^{(p)}_{a,{\rm tree}}(\bfk_1,\cdots,\bfk_p;\eta)&=
e^{p\,\eta}\,F_{a}^{(p)}(\bfk_1,c\dots,\bfk_p),
\label{eq:Gamma-p_tree}
\end{align}
for the tree-level contribution, and 
\begin{widetext}
\begin{align}
\Gamma_{a,n\mbox{-}{\rm loop}}^{(p)}
(\bfk_1,\cdots,\bfk_p;\eta)
&=e^{(2n+p)\,\eta}\,c_n^{(p)}
\int\frac{d^3\bfp_1\cdots d^3\bfp_n}{(2\pi)^{3n}}\,
F_a^{(2n+p)}(\bfp_1,-\bfp_1,\cdots,\bfp_n,-\bfp_n,\bfk_1,\cdots,\bfk_p)
P_0(p_1)\cdots P_0(p_n)
\nonumber\\
&\equiv e^{(2n+p)\,\eta}\,\,
\overline{\Gamma}_{a,n\mbox{-}{\rm loop}}^{(p)}(\bfk_1,\cdots,\bfk_p)
\label{eq:Gamma-p_nloop}
\end{align}
\end{widetext}
for the $n$-loop order contributions, with the coefficient $c_n^{(p)}$ 
given by 
\begin{align}
& c_n^{(p)}=\,
\left( 
\begin{array}{c}
2n+p
\\
p
\end{array}
\right) \ (2n-1)!!. 
\end{align}
Note that $n$-loop order correction $\Gamma^{(p)}_{a,n\mbox{-}{\rm loop}}$ 
is that each perturbative correction possesses the following asymptotic 
form,
\begin{align}
\Gamma^{(p)}_{a,n\mbox{-}{\rm loop}}
\stackrel{k\to\infty}{\longrightarrow} 
\frac{1}{n!} \left(-\frac{k^2\sigmav^2e^{2\eta}}{2}\right)^{n}
\Gamma^{(p)}_{a,n\mbox{-}{\rm tree}}, 
\label{eq:Gamma}
\end{align}
which consistently recovers the expression (\ref{eq:Gamma_high-k}) 
when we sum up all the loop contributions. This fact leads to a 
novel regularized scheme, in which the low- and high-$k$ behaviors are 
smoothly interpolated without any ambiguities \cite{Bernardeau:2011dp}. 
Then, the {\it regularized} propagators are expressed in a transparent way 
in terms of the standard PT results, and one gets 
\begin{widetext}
\begin{align}
\Gamma_{a,{\rm reg}}^{(p)}(\bfk_1,\cdots,\bfk_p;\eta)=e^{p\,\eta}\left[
F^{(p)}_a(\bfk_1,\cdots,\bfk_p)\left\{1+
\frac{k^2\sigmav^2e^{2\eta}}{2}\right\}+
e^{2\eta}\,\overline{\Gamma}^{(p)}_{a,{\rm 1\mbox{-}loop}}(\bfk_1,
\cdots,\bfk_p)\right]
\exp\left\{-\frac{k^2\sigmav^2e^{2\eta}}{2}\right\}, 
\end{align}
\end{widetext}
which consistently reproduces one-loop PT results at low-$k$. 
This construction is easily generalized to include the two-loop order
PT corrections at low-$k$: 
\begin{widetext}
\begin{align}
&\Gamma_{a,{\rm reg}}^{(p)}(\bfk_1,\cdots,\bfk_p;\eta)=e^{p\,\eta}\left[
F^{(p)}_a(\bfk_1,\cdots,\bfk_p)\left\{1+
\frac{k^2\sigmav^2e^{2\eta}}{2}+
\frac{1}{2}\left(\frac{k^2\sigmav^2e^{2\eta}}{2}\right)^2\right\}
\right.
\nonumber\\
&\left.\qquad\qquad\qquad+\,
e^{2\eta}\,\overline{\Gamma}^{(p)}_{\rm 1\mbox{-}loop}(\bfk_1,\cdots,\bfk_p)
\left\{1+\frac{k^2\sigmav^2e^{2\eta}}{2}\right\}+e^{4\eta}\,
\overline{\Gamma}^{(p)}_{\rm 2\mbox{-}loop}(\bfk_1,\cdots,\bfk_p)\right]
\exp\left\{-\frac{k^2\sigmav^2e^{2\eta}}{2}\right\}.
\end{align}
\end{widetext}
Note that the functions $\overline{\Gamma}^{(p)}_{n\mbox{-}{\rm loop}}$ are the 
scale-dependent part of the propagator defined 
by Eq.~(\ref{eq:Gamma-p_nloop}).

\subsection{Redshift-space power spectrum from regularized $\Gamma$ expansion}
\label{subsec:pk_from_RegPT}

In what follows, with the regularized prescription of the 
multi-point propagators, we apply
the $\Gamma$ expansion to calculate the redshift-space power spectrum and 
correlation function up to the two-loop order. We hereafter call 
the PT treatment with regularized $\Gamma$ expansion RegPT. 
Here, we briefly describe the technical implementation 
of the RegPT to the model of RSD. 
Readers who are only interested in the results may skip this subsection 
and directly go to next section. 

In Eq.(\ref{eq:TNS_model}) ignoring the $D$ term, there appears
three terms to be computed perturbatively, i.e., $P_{\rm Kaiser}$, $A$ 
and $B$ terms, which include the real-space power spectrum and bispectrum. 
Below, we will separately give a prescription how to compute each term.

\subsubsection{Nonlinear Kaiser term $P_{\rm Kaiser}$}
\label{subsubsec:Kaiser_term}

The power spectrum $P_{\rm Kaiser}$ consists of the three real-space power 
spectra, $\Pdd$, $\Pdv$, and $\Pvv$. 
In terms of the power spectra $P_{ab}$ for the doublet $\Psi_a$, 
these are equivalent to $P_{11}$, $P_{12}$, and $P_{22}$, respectively. 
Thus, for a practical computation of $P_{\rm Kaiser}$, we just follow the 
prescription presented in Ref.~\cite{Taruya:2012ut}, and 
use the same technique to calculate each power spectrum contribution 
at one- and two-loo order. In Appendix \ref{app:Gamma_expansion_Pk_Bk}, 
we give explicit expressions 
for the power spectra $P_{ab}$ up to the two-loop order.

\subsubsection{$A$ term}
\label{subsubsec:A_term}

The $A$ term given in Eq.~(\ref{eq:A_term}) 
includes the cross bispectrum $B_\sigma$ [Eq.~(\ref{eq:def_B_sigma})], 
and we thus need the RegPT to explicitly compute this term. 
The function $B_\sigma$ is expressed in terms of the bispectra of $\Psi_a$: 
\begin{align}
 B_\sigma(\bfk_1,\bfk_2,\bfk_3) &=  B_{211}(\bfk_1,\bfk_2,\bfk_3)
\nonumber\\
& +f\left(\frac{k_{2,z}}{k_2}\right)^2\,B_{221}(\bfk_1,\bfk_2,\bfk_3)
\nonumber\\
&+f\left(\frac{k_{3,z}}{k_3}\right)^2\,B_{212}(\bfk_1,\bfk_2,\bfk_3)
\nonumber\\
&+f^2\left(\frac{k_{2,z}k_{3,z}}{k_2k_3}\right)^2\,B_{222}(\bfk_1,\bfk_2,\bfk_3).
\end{align}
Note that the $A$ term itself is already a higher-order contribution, and 
in computing the redshift-space power spectrum, 
the tree-level and one-loop calculations of the bispectrum are sufficient 
for a consistent calculation of $P^{\rm(S)}$ up to the 
one- and two-loop order, respectively. The expressions for 
the regularized bispectra are explicitly given in Appendix 
\ref{app:Gamma_expansion_Pk_Bk}.

In computing the $A$ term, the expression given in Eq.~(\ref{eq:A_term}) 
is not suited for a practical purpose. Here, following the 
same technique as used in Ref.~\cite{Taruya:2012ut}, we derive alternative
expression for which the term is expanded as the polynomials of 
$\mu$ and $f$. The detail of derivation is described in  
 Appendices A and B of Ref.~\cite{Taruya:2010mx}, and we here 
present the final expression, in which the three-dimensional 
integral is reduced to the sum of the two-dimensional integrals:   
\begin{widetext}
\begin{align}
& A(k,\mu) 
=\sum_{n=1}^3\sum_{a,b=1}^2\,\mu^{2n}\,f^{a+b-1}\,
\frac{k^3}{(2\pi)^2}\,\int_0^\infty dr\int_{-1}^{1} dx
\,\left\{A^n_{ab}(r,x)\,B_{2ab}(\bfp,\bfk-\bfp,-\bfk)+
\widetilde{A}^n_{ab}(r,x)\,B_{2ab}(\bfk-\bfp,\bfp,-\bfk)
\right\}.  
\label{eq:A_term_simplified}
\end{align}
\end{widetext}
Here, $r$ and $x$ are the dimensionless variables 
defined by $r=p/k$, and $x=(\bfk\cdot\bfp)/(kp)$. 
The non-vanishing components of $A^a_{bc}$ and $\widetilde{A}^a_{bc}$ 
are summarized as follows:
\begin{align}
& A^1_{11}=r\,x, \quad
A^1_{21}=-\frac{r^2(-2+3rx)(x^2-1)}{2(1+r^2-2rx)},
\nonumber\\
& A^2_{21}=\frac{r\{2x+r(2-6x^2)+r^2x(-3+5x^2)\}}{2(1+r^2-2rx)},
\nonumber\\
& A^2_{12}=A^1_{11},\quad
A^2_{22}=A^1_{21},\quad
A^3_{22}=A^2_{21},
\nonumber\\
& \widetilde{A}^1_{11}=-\frac{r^2(rx-1)}{1+r^2-2rx},\quad
\widetilde{A}^1_{21}=\frac{r^2(-1+3rx)(x^2-1)}{2(1+r^2-2rx)},
\nonumber\\
& \widetilde{A}^2_{21}=-\frac{r^2\{1-3x^2+rx(-3+5x^2)\}}{2(1+r^2-2rx)},
\nonumber\\
& \widetilde{A}^2_{12}=\widetilde{A}^1_{11},\quad
\widetilde{A}^2_{22}=\widetilde{A}^1_{21},\quad
\widetilde{A}^3_{22}=\widetilde{A}^2_{21}.
\nonumber
\end{align}

\begin{table*}[ht]
\caption{\label{tab:nbody_nishimichi} Cosmological parameters for $N$-body 
simulations ($\Lambda$CDM) \cite{Taruya:2012ut}}
\begin{ruledtabular}
\begin{tabular}{lcccc|ccccccc}
Name & $L_{\rm box}$ & \# of particles & $z_{\rm ini}$ 
& \# of runs & 
$\Omega_{\rm m}$ & $\Omega_{\Lambda}$ & $\Omega_{\rm b}/\Omega_{\rm m}$ & $w$ & $h$ & $n_s$ & $\sigma_8$ \\
\hline
\verb|wmap5| & $2,048h^{-1}$Mpc & $1,024^3$ & $15$ & $60$ & 
0.279 &  0.721 & 0.165 & -1 & 0.701 & 0.96 & $0.815_9$\\
\end{tabular}
\end{ruledtabular}
\end{table*}

\subsubsection{$B$ term}
\label{subsubsec:B_term}

The expression of the $B$ term in Eq.~(\ref{eq:B_term}) 
involves the integral of the power spectra, to which we 
apply the RegPT. Note that 
the $B$ term itself is already higher-order contribution, 
and for a consistent calculation of the redshift-space power spectrum,  
the tree-level and one-loop calculation of the real-space power spectra are
sufficient.

In Ref.~\cite{Taruya:2010mx}, the expression of the $B$ term suited for 
a practical calculation
has been derived without employing the perturbative 
calculations: 
\begin{widetext}
\begin{eqnarray}
B(k,\mu)=\sum_{n=1}^4\,\, \sum_{a,b=1}^2 \mu^{2n}(-f)^{a+b} 
\frac{k^3}{(2\pi)^2}\int_0^\infty dr \int_{-1}^{+1} dx 
\,B^n_{ab}(r,x)\,\frac{P_{a2}\left(k\sqrt{1+r^2-2rx}\right)P_{b2}(kr)}
{(1+r^2-2rx)^a}, 
\label{eq:B_PT_formula}
\end{eqnarray}
\end{widetext}
The coefficients $B^n_{ab}$ are given in Appendix B of 
Ref.~\cite{Taruya:2010mx}.

\section{Results}
\label{sec:results}

In this section, we present the results of the PT calculations, and compare 
the PT predictions with $N$-body simulations. 
The redshift-space power spectrum and correlation function are computed with 
the RegPT 
consistently including the PT corrections up to the one- and two-loop 
orders. Note that for the one- and two-loop predictions, 
the bispectra and power spectra in the $A$ and $B$ term have been consistently 
computed including the PT corrections up to the tree- and one-loop orders, 
respectively. After briefly describing the $N$-body simulations in 
Sec.~\ref{subsec:simulation}, the comparison between the predictions and 
simulations is presented for power spectrum 
in Sec.~\ref{subsec:pkred} and for correlation function in 
Sec.~\ref{subsec:xired}.

\subsection{$N$-body simulations}
\label{subsec:simulation}

To compare the model prediction with $N$-body simulations, 
we use the data set presented in our previous paper \cite{Taruya:2012ut}.
The data were created by a public $N$-body code \verb|GADGET2| 
\cite{Springel:2005mi} 
with cubic boxes of side length $2,048\,h^{-1}$Mpc, and $1,024^3$ particles. 
The cosmological parameters adopted in these $N$-body simulations are 
basically the same as in the previous one, and are determined 
by the five-year WMAP observations \cite{Komatsu:2008hk} 
(see Table~\ref{tab:nbody_nishimichi}). 
The initial conditions were generated by the \verb|2LPT| code 
\cite{Crocce:2006ve} with the initial redshift $z_{\rm init}=15$, and  
the results of $60$ independent realizations 
are stored at redshifts $z=3$, $2$, $1$, and $0.35$. The 
total volume of each output is $515\,h^{-3}$Gpc$^3$.

We measure both the matter power spectrum and 
correlation function in redshift space, 
applying the distant-observor approximation. For the power spectrum, 
we adopt 
the Cloud-in-Cells interpolation, and construct 
the Fourier transform of the density field assigned on the $1,024^3$ grids. 
As for the estimation of two-point 
correlation function, we adopt the grid-based calculation using the 
Fast Fourier Transformation \cite{Taruya:2009ir}. Similar to the 
power spectrum analysis, we first compute the square of the density 
field on each grid of Fourier space. Then, applying the inverse Fourier 
transformation, we take the average over realization, 
and finally obtain the two-point correlation function. The 
implementation of this method, together with the convergence test, is 
presented in more detail in Ref.~\cite{Taruya:2009ir}. 
In what follows, the error bars of the $N$-body results 
indicate the root-mean-square fluctuations of the averaged 
power spectra or correlation functions over the 60 realizations.

\begin{figure}[ht]

\vspace*{-2cm}

\begin{center}
\includegraphics[width=8.5cm,angle=0]{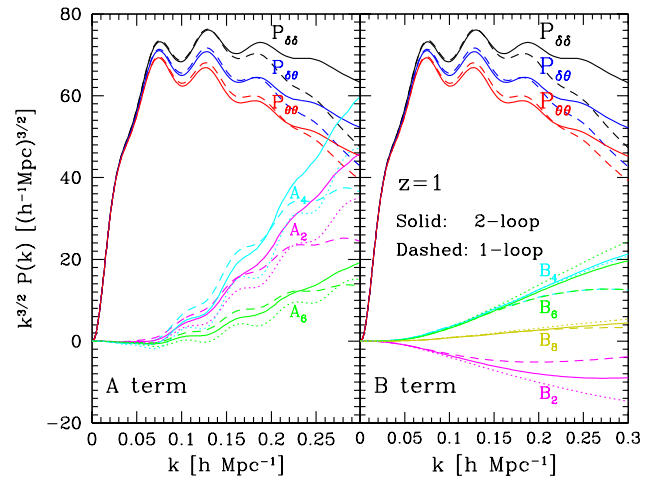}
\end{center}

\vspace*{-0.3cm}

\caption{Contribution of the correction terms for the PT model of 
redshift-space power spectrum. Left and right panels respectively show 
the $A$ and $B$ terms at $z=1$, respectively. 
For illustrative purpose, all the results are multiplied by $k^{3/2}$. 
The $A$ and $B$ terms are expanded as 
$A(k,\mu)=\sum_n^3 \mu^{2n}\,A_{2n}(k)$ and  
$B(k,\mu)=\sum_n^4 \mu^{2n}\,B_{2n}(k)$.  
Here, we plot 
the scale-dependent coefficients $A_{2n}$ and $B_{2n}$ 
($A_2,\,\,B_2$: magenta, $A_4,\,\,B_4$: cyan, $A_6\,\,B_6$: green, $B_8$: 
yellow). In each panel, 
solid lines are the corrections for the two-loop contributions, 
while the dashed lines are the results for the one-loop contributions.  
For references, the power spectra $P_{\delta\delta}$, $P_{\delta\theta}$, and 
$P_{\theta\theta}$ computed from RegPT 
are also shown in black, blue and red lines, respectively.  
\label{fig:pkcorr_A_B}}
\end{figure}

\subsection{Power spectrum}
\label{subsec:pkred}

\begin{figure*}
\begin{center}
\hspace*{-0.7cm}
\includegraphics[width=6.3cm,angle=0]{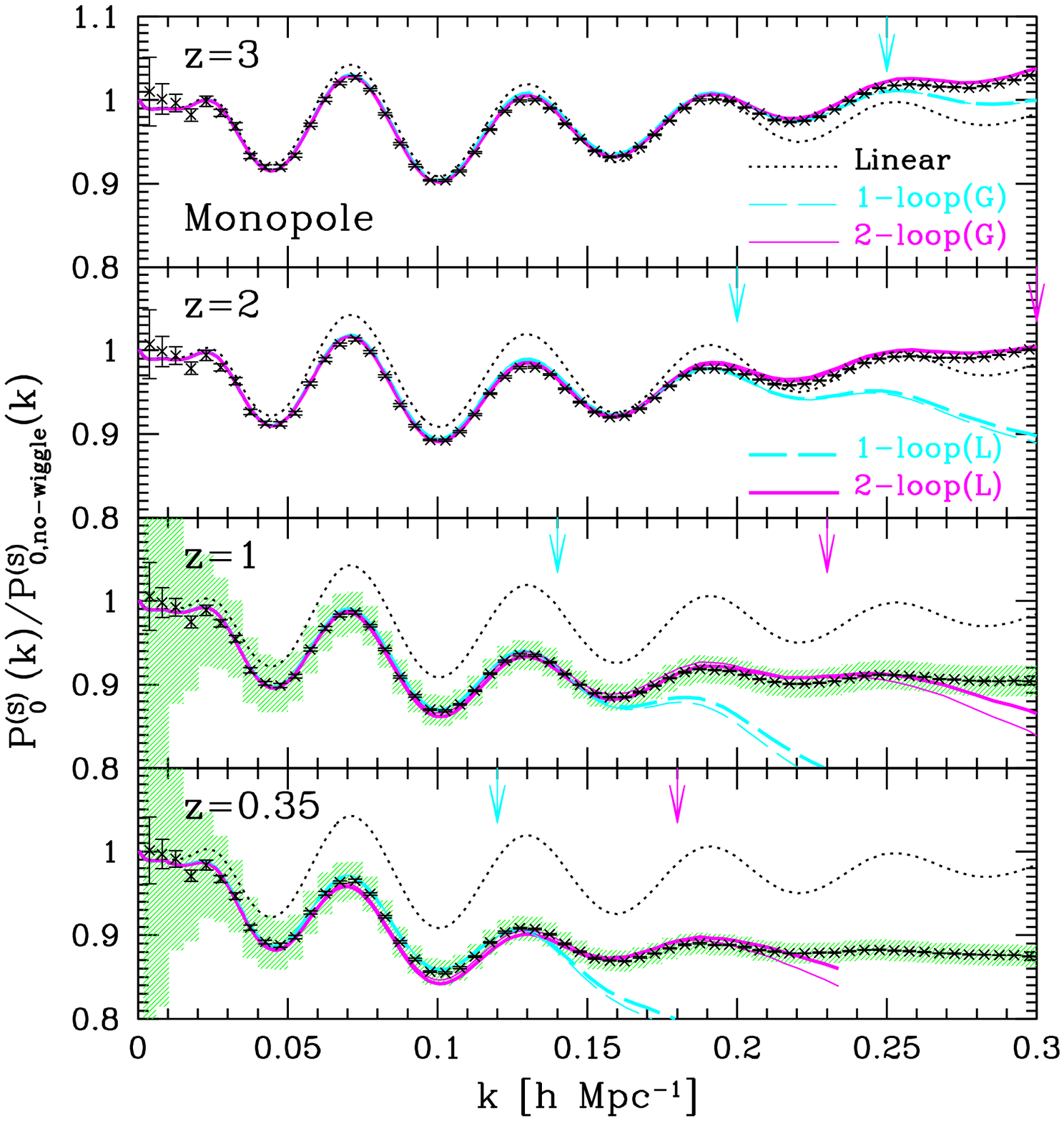}
\hspace*{-0.5cm}
\includegraphics[width=6.3cm,angle=0]{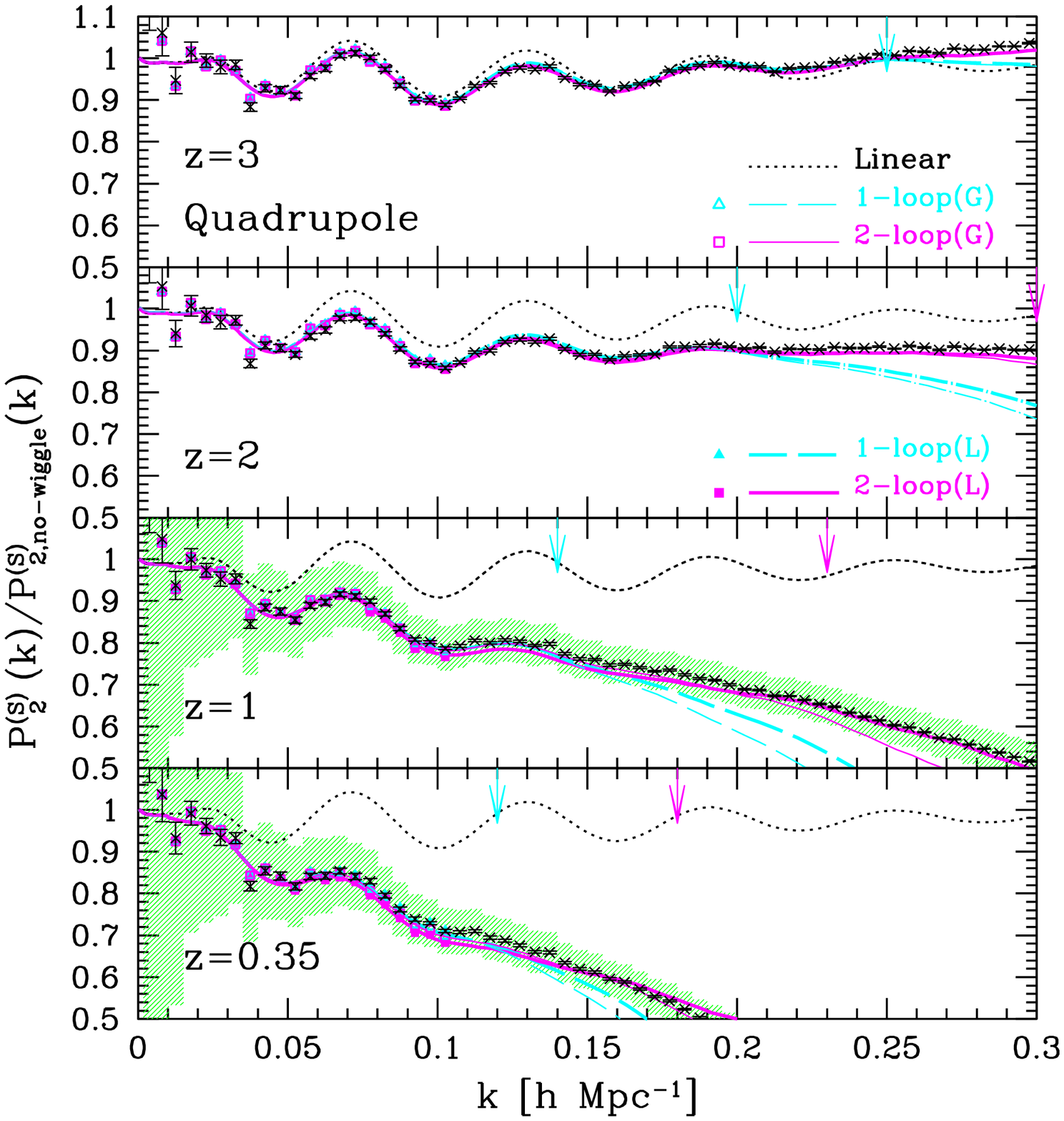}
\hspace*{-0.5cm}
\includegraphics[width=6.3cm,angle=0]{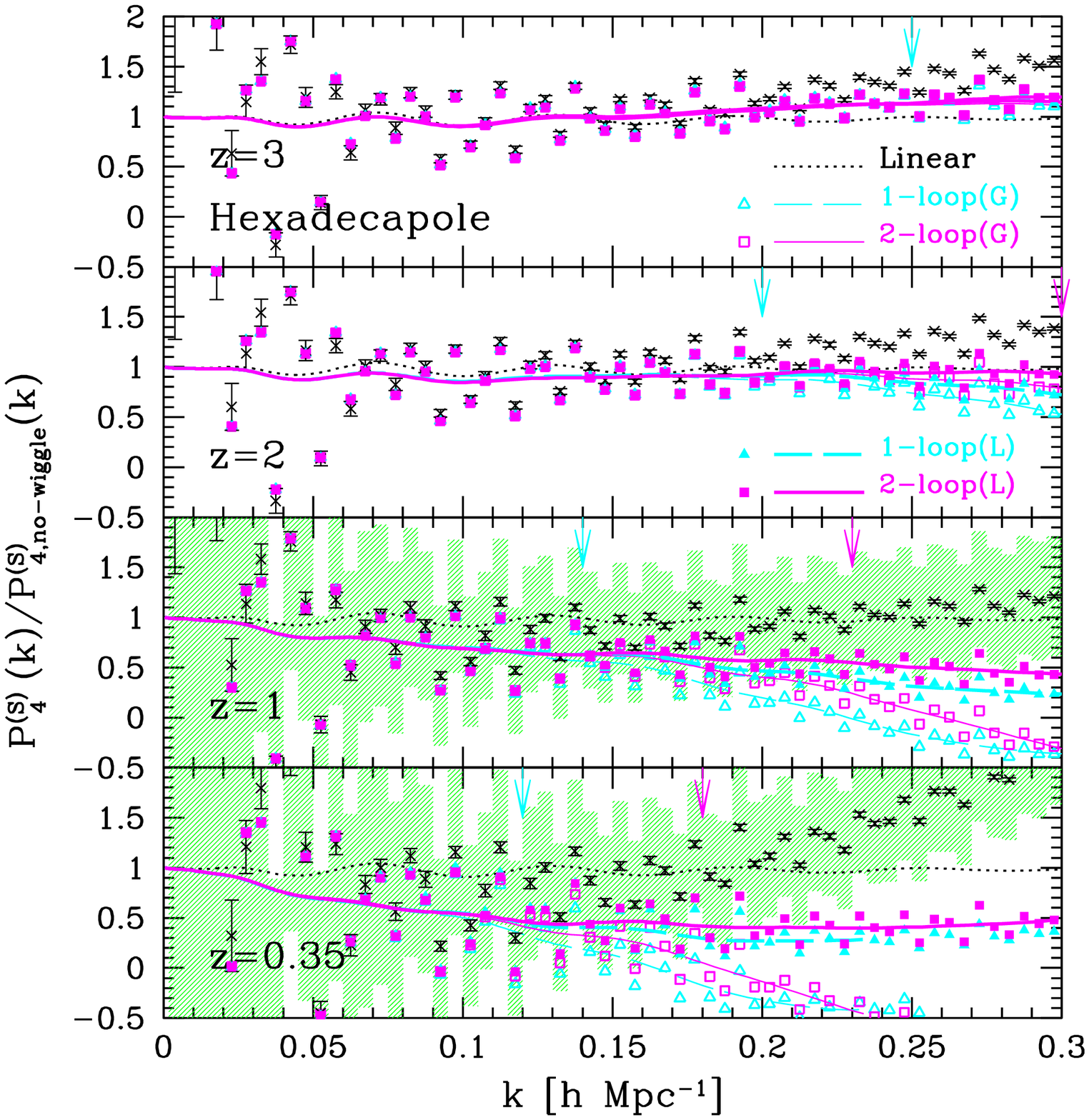}
\end{center}

\vspace*{-0.5cm}

\caption{Ratio of power spectra to the smoothed reference spectra in 
  redshift space, 
  $P_\ell^{\rm (S)}(k)/P_{\ell,{\rm no\mbox{-}wiggle}}^{\rm (S)}(k)$. Left, 
  middle, and right panels respectively show the monopole ($\ell=0$), 
  quadrupole ($\ell=2$) and hexadecapole ($\ell=4$) 
  contributions to the redshift-space power spectrum. N-body 
  results are taken from the {\tt wmap5} simulations of 
  Ref.~\cite{Taruya:2012ut}. The reference spectrum 
  $P_{\ell,{\rm no\mbox{-}wiggle}}^{\rm (S)}$ is computed with 
  the no-wiggle approximation of the linear transfer function 
  \cite{Eisenstein:1997ik}, taking
  account of the linear theory of the Kaiser effect. 
  Long-dashed and solid lines respectively indicate the results based on 
  the RegPT calculations at one- and two-loop orders,  
  adopting the Gaussian (thin) and Lorentzian (thick) form of the 
  damping functions. 
  Triangles and squares in middle and right panels are also obtained 
  from the same calculation at one- and two-loop orders, 
  but taking account of the effect of finite grid-size for the power 
  spectrum measurement in $N$-body simulations (see text and Appendix 
  \ref{App:finite_grid} in detail). For comparison, 
  the $1$-$\sigma$ statistical errors of the hypothetical survey with volumes 
  $V=5\,h^{-3}$\,Gpc$^{3}$ and number density 
  $n=5\times10^{-4}\,h^3$\,Mpc$^{-3}$ are estimated from 
  Eq.~(\ref{eq:error_pk}), and are depicted as green shaded regions 
  around the N-body results at $z=0.35$ and $1$. 
\label{fig:ratio_pk_red_PT}}
\end{figure*}

Before presenting a quantitative comparison, we 
first look at the contribution of each term appeared in the model of RSD. 
In Fig.~\ref{fig:pkcorr_A_B}, for specific redshift $z=1$, 
we plot the results for $A$ and $B$ terms as well as the 
power spectra $\Pdd$, $\Pdv$, and $\Pvv$. From the expressions 
given in Eqs.~(\ref{eq:A_term_simplified}) and (\ref{eq:B_PT_formula}), 
the $A$ and $B$ term can be expanded as 
$A(k,\mu)=\sum_n^3 A_{2n}(k)\mu^{2n}$ and 
$B(k,\mu)=\sum_n^4 B_{2n}(k)\mu^{2n}$, and we here plot the 
scale-dependent coefficients $A_{2n}$ and $B_{2n}$ multiplied by $k^{3/2}$ 
($A_2,\,\,B_2$: magenta, $A_4,\,\,B_4$: cyan, $A_6,\,\,B_6$: green, $B_8$: 
yellow). Dashed and solid lines respectively indicate 
the one- and two-loop contributions to the redshift-space power spectrum. 
For reference, we also plot the results with standard PT calculations 
in dotted lines. Compared to the standard PT 
results, the coefficients of the $A$ term 
are slightly enhanced at the two-loop order,  and the oscillatory 
feature originating from the BAOs is somewhat smeared. This is 
similar to what we saw in the real-space power spectrum. 
Fig.~\ref{fig:pkcorr_A_B} apparently indicates that 
at the two-loop order, the $A$ term seem to eventually dominate the total 
power spectrum at small scales. However, this is actually not true. 
Because of the exponential cutoff generically appeared in the 
multi-point propagators, 
the amplitudes of both the $A$ and $B$ terms are suppressed 
at small scales, as similarly seen in the power spectra of the density and 
velocity fields. This regularized UV property enables us to give a 
convergent result for the correlation function,   
although, as a trade-off, 
the prediction of redshift-space power spectrum eventually 
becomes inappropriate at some small scales.

\begin{figure}

\vspace*{-1cm}

\begin{center}
\includegraphics[width=8.5cm,angle=0]{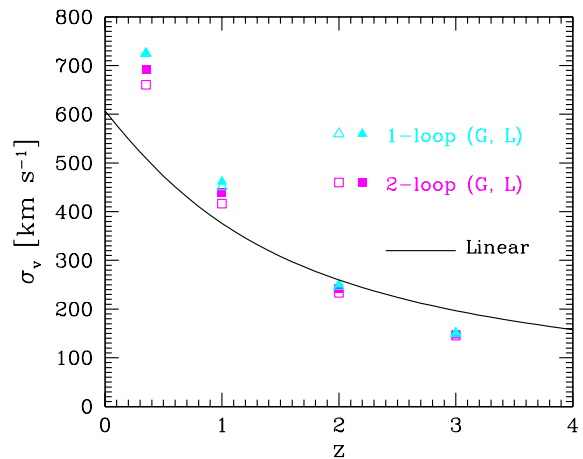}
\end{center}

\vspace*{-2cm}

\caption{Fitted values of  $\sigma_{\rm v}$ as function redshift. 
Triangles and squares are respectively obtained by fitting the 
one- and two-loop PT predictions of the monopole and 
quadrupole spectra to those obtained from 
the $N$-body simulations. Open (filled) symbols are the fitting results 
adopting the Gaussian (Lorentzian) form of the damping function 
in the theoretical predicitons [see Eq.~(\ref{eq:Damping_func})]. 
For reference, 
we also plot the linear theory prediction in solid line. 
\label{fig:sigmav2_fit}}
\end{figure}

Now, let us compare the predictions of redshift-space power spectrum 
with those of the $N$-body simulations in greater detail, and 
investigate the extent to which the PT model reproduces the simulation 
results. Fig.~\ref{fig:ratio_pk_red_PT} plots the ratio of power spectra to 
the smooth reference spectra, 
$P_\ell^{\rm(S)}(k)/P^{\rm(S)}_{\ell,{\rm no\mbox{-}wiggle}}(k)$, 
where $P^{\rm(S)}_{\ell,{\rm no\mbox{-}wiggle}}(k)$ is the 
linear power spectrum computed with the no-wiggle formula of 
Ref.~\cite{Eisenstein:1997ik}.  
The power spectrum $P^{\rm(S)}_\ell$ is the $\ell$-th order moment of 
the redshift-space power spectrum defined by
\begin{align}
P_\ell^{\rm(S)}(k)=\frac{2\ell+1}{2}\,\int_{-1}^1 d\mu\,P^{\rm(S)}(k,\mu)
\,\mathcal{P}_\ell(\mu),
\end{align}
with $\mathcal{P}_\ell$ being the Legendre polynomials.
In Fig.~\ref{fig:ratio_pk_red_PT},  
the results of the monopole ($\ell=0$), quadrupole ($\ell=2$), 
and hexadecapole ($\ell=4$) power spectra are 
respectively shown from left to right panels. Here, the
dashed and solid lines are the results
based on the RegPT calculations at the one- and two-loop orders. 
In plotting these predictions, the velocity dispersion $\sigmav$ in 
Eq.~(\ref{eq:TNS_model}) is treated as a free parameter, and is 
determined by fitting the 
model prediction to the $N$-body results of monopole and quadrupole spectra,  
assuming the Gaussian (thin) and Lorentzian (thick) 
forms of the damping function $D_{\rm FoG}$:
\begin{align}
D_{\rm FoG}(x)= \left\{
\begin{array}{ccl}
e^{-x^2}      &;& \mbox{Gaussian}\\
1/(1+x^2/2)^2&;& \mbox{Lorentzian}
\end{array}
\right.
\label{eq:Damping_func}
\end{align}
The predictions were fitted to the simulation results up to the 
scale indicated by the vertical arrows. Note that this roughly corresponds to 
the critical wavenumber $k_{\rm crit}$, 
below which the RegPT prediction in real space agrees 
with $N$-body simulation at a percent-level precision \cite{Taruya:2012ut}. 
The fitted values of $\sigmav$ are summarized in Fig.~\ref{fig:sigmav2_fit}.

Overall, the agreement between predictions and simulations is remarkably 
good for the monopole and quadrupole spectra. In both one- and two-loop
results, the range of agreement is almost comparable to what we found in 
the real-space comparison. This is true irrespective of the choice of 
the damping function. In particular, the two-loop results look very similar 
to what we obtained with
the closure \cite{Taruya:2007xy,Taruya:2009ir} 
and standard PT calculations \cite{Taruya:2010mx}. Rather, with 
the RegPT, the oscillatory feature in $A$ term is erased, 
and the contribution of the $A$ term to the BAO structure is somewhat 
reduced. As a result, the total sum of each contribution closely 
matches the $N$-body results better than the previous results.

Turning to the hexadecapole power spectra, on the other hand, 
the simulation results show somewhat noisy behaviors, and it is 
bit difficult to compare those with the predictions depicted as 
continuous lines. This noisy structure basically comes from 
the fact that the power spectra are measured from the grid-assigned 
density field. Because of 
the oscillatory feature of the Legendre polynomials $\mathcal{P}_\ell$,   
the measurement of the higher-multipole spectra tends to be sensitively 
affected by the finite grid-size. In order to remedy this, 
we also take account of the effect of finite grid-size in the PT prediction, 
and compute the hexadecapole spectra in the same way as we did 
in the $N$-body simulations. The detailed prescription of this treatment 
is presented in Appendix \ref{App:finite_grid}.

The predictions taking account of the finite-grid size effect 
are depicted as the triangles and squares for one- and two-loop calculations, 
respectively. At high-$z$, the PT results faithfully reproduces 
the noisy behavior in the simulation results. In particular, 
at low-$k$, the symbols almost overlap each other, indicating that the 
simulations consistently recover the linear theory prediction. 
At lower redshifts, on the other hand, there appear small but 
non-negligible discrepancies. To see the significance of this, we 
consider the hypothetical galaxy survey, and estimate 
the expected $1$-$\sigma$ statistical errors, $\Delta P_\ell^{\rm(S)}$, 
depicted as green shaded region at $z=0.35$ and $1$ 
in Fig.~\ref{fig:ratio_pk_red_PT}. The statistical error 
$\Delta P_\ell^{\rm(S)}$ is simply computed 
with 
\begin{align}
[\Delta P_\ell^{\rm(S)}(k)]^2=\frac{2}{N_k} \,\sigma_{P,\ell}^2(k)
\,\,;\quad N_k=\frac{4\pi\,k^2\Delta k}{(2\pi/V^{1/3})^3},
\label{eq:error_pk}
\end{align}
with the function $\sigma_{P,\ell}$ given by 
\begin{align}
\sigma_{P,\ell}^2(k)=
\frac{(2\ell+1)^2}{2}\int_{-1}^1d\mu\left\{
P^{\rm(S)}_{\rm lin}(k,\mu)+\frac{1}{n}\right\}^2,  
\label{eq:sigma_P}
\end{align}
where $P^{\rm(S)}_{\rm lin}$ is the linear power spectrum, and the 
$\Delta k$ is the bin width for which we adopt the same bin size used in 
the $N$-body data.  
The volume and the number density of the hypothetical survey are respectively 
set to $V=5\,h^{-3}$\,Gpc$^3$ and $n=5\times10^{-4}\,h^3$\,
Mpc$^{-3}$, which roughly correspond to those of the 
Baryon Oscillation Spectroscopic Survey (BOSS) 
\footnote{{\tt http://www.sdss3.org}} 
or the survey with Subaru Measurement of Images and Redshifts (SuMIRe) with 
Prime Focus Spectrograph (PFS) 
\footnote{{\tt http://sumire.ipmu.jp/en/}} 
\footnote{Strictly speaking, the BOSS aims 
at observing galaxies at $0.2\lesssim z\lesssim0.8$, while 
SuMIRe PFS project will observe galaxies at $0.6\lesssim z\lesssim2.4$ 
\cite{Ellis:2012rn}}. 
Then, we found that the predicted 
monopole and quadrupole spectra agree with simulations well within 
the statistical error, while the discrepancy in  
the hexadecapole spectra is marginal or even exceeds the $1$-$\sigma$ error 
at high-$k$, depending on the functional form of the damping function 
$D_{\rm FoG}$. 
These results not only indicate the sensitive dependence of the damping 
function but also suggest a possible deficit in the RSD model when 
predicting the higher-multipole spectra. This point will be further discussed 
in greater detail in Sec.~\ref{sec:discussion}.

\subsection{Correlation function}
\label{subsec:xired}

\begin{figure*}
\begin{center}
\hspace*{-0.7cm}
\includegraphics[width=6.7cm,angle=0]{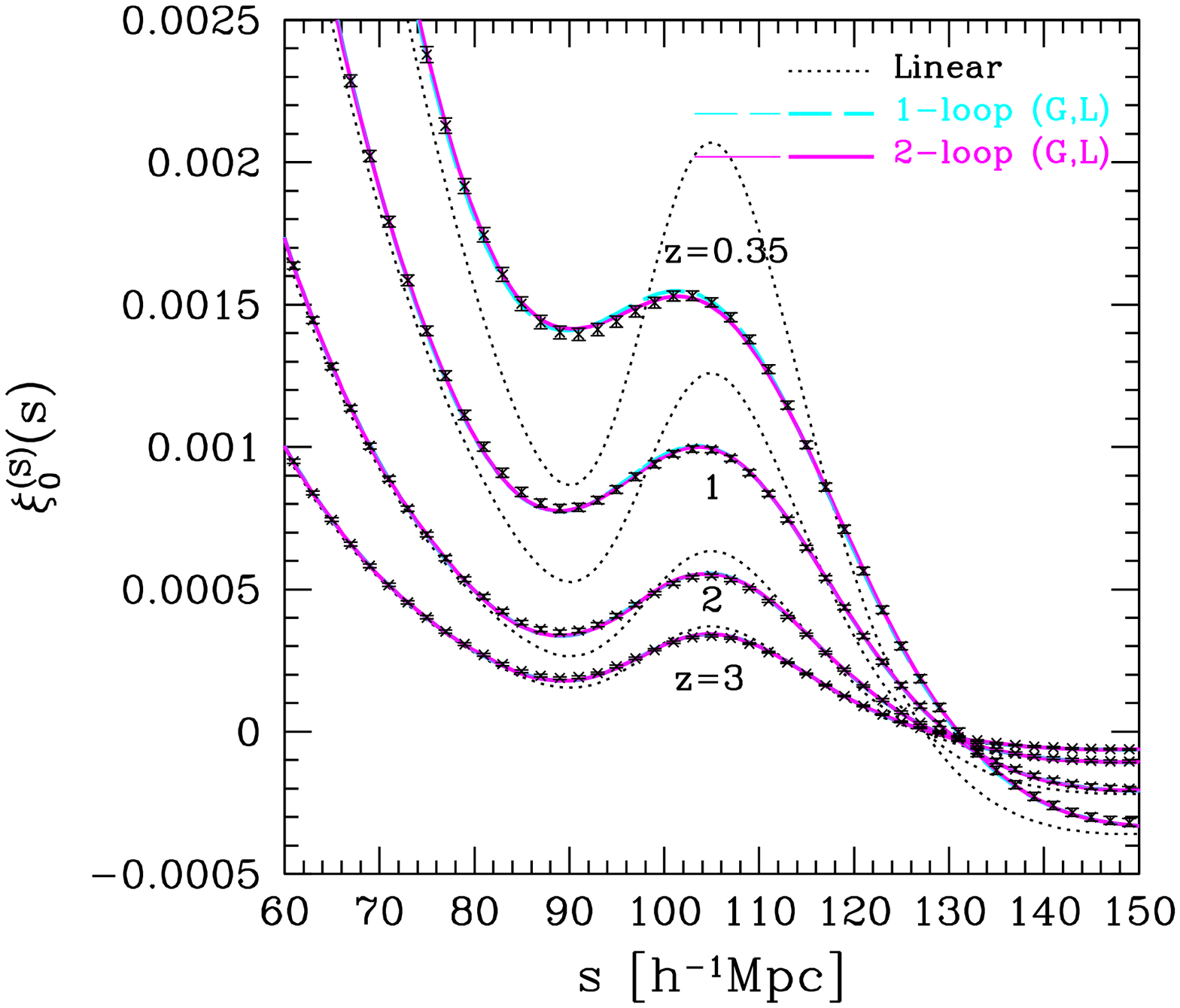}
\hspace*{-1.0cm}
\includegraphics[width=6.7cm,angle=0]{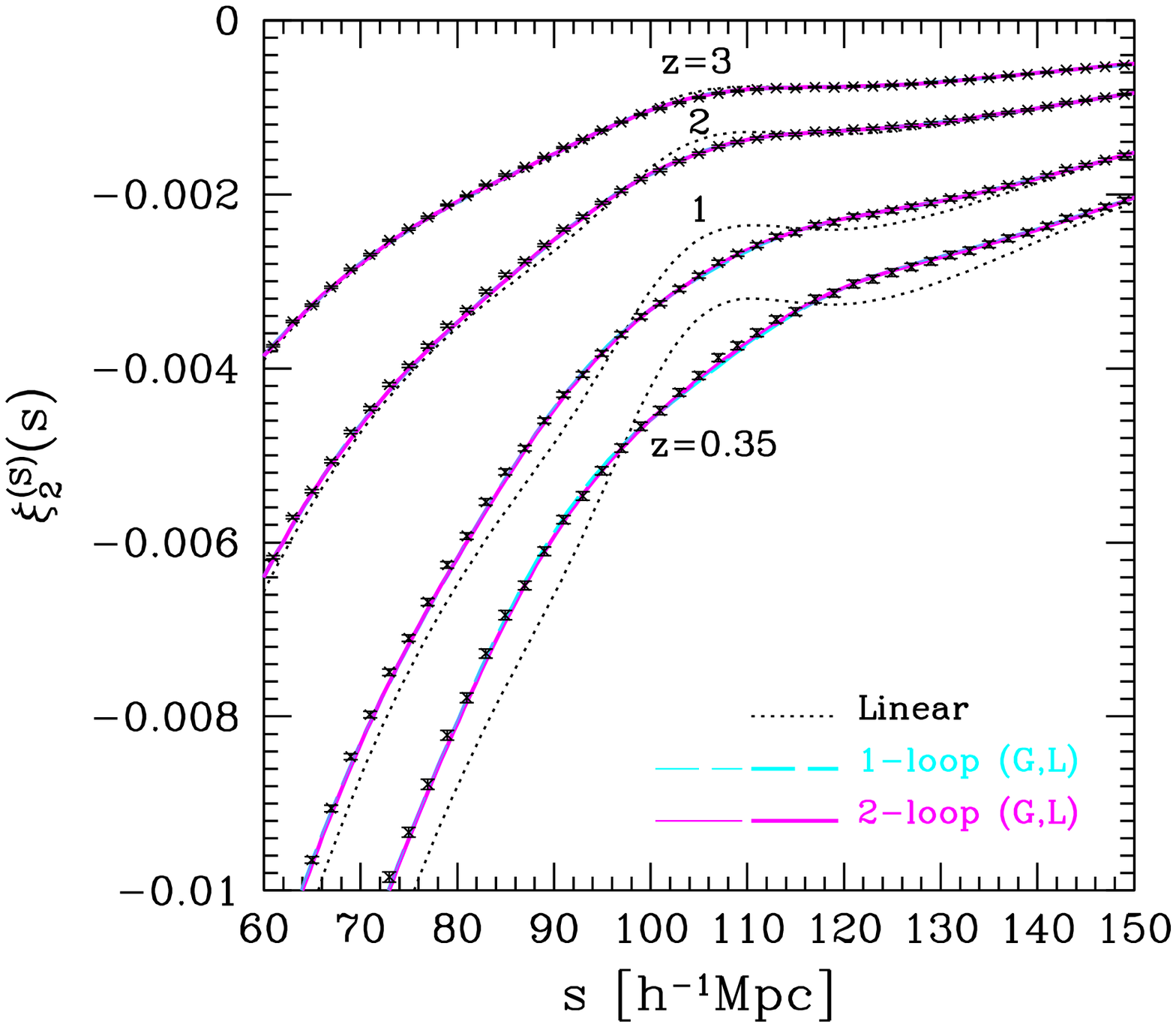}
\hspace*{-1.0cm}
\includegraphics[width=6.7cm,angle=0]{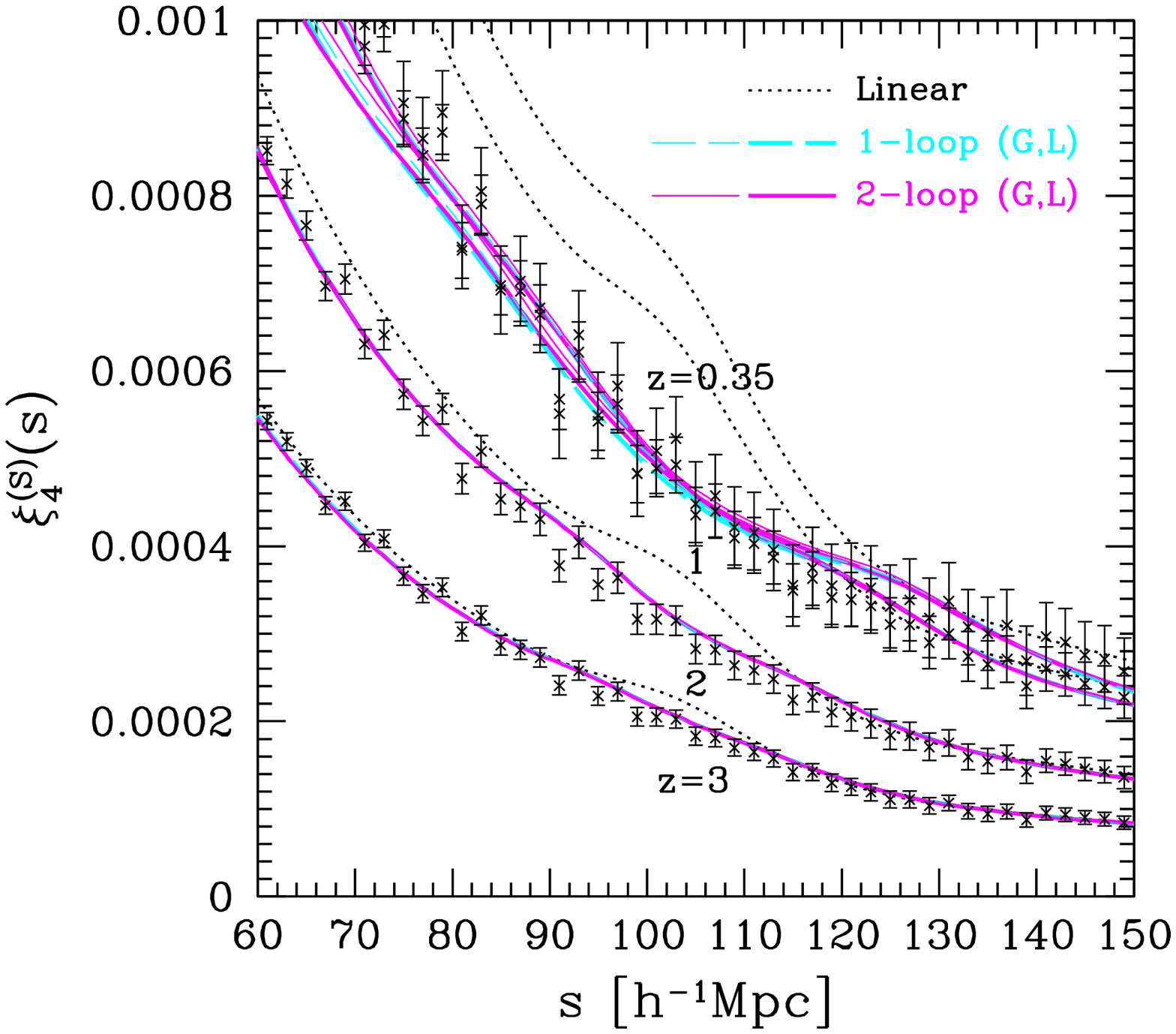}
\end{center}

\vspace*{-1.0cm}

\caption{Redshift-space correlation functions around the baryon acoustic peak. 
Monopole ($\ell=0$), quadrupole ($\ell=2$), and hexadecapole ($\ell=4$) 
moments of the redshift-space correlation function 
are respectively shown in left, middle and right panels. Dotted lines 
are the linear theory predictions, while the dashed and solid lines 
respectively represent the results based on the RSD model 
using the RegPT up to the one- and two-loop orders, adopting the 
Gaussian (thin) and Lorentzian (thick) damping function. 
\label{fig:xi_red_PT1}}
\end{figure*}

\begin{figure*}
\begin{center}
\hspace*{-0.9cm}
\includegraphics[width=6.6cm,angle=0]{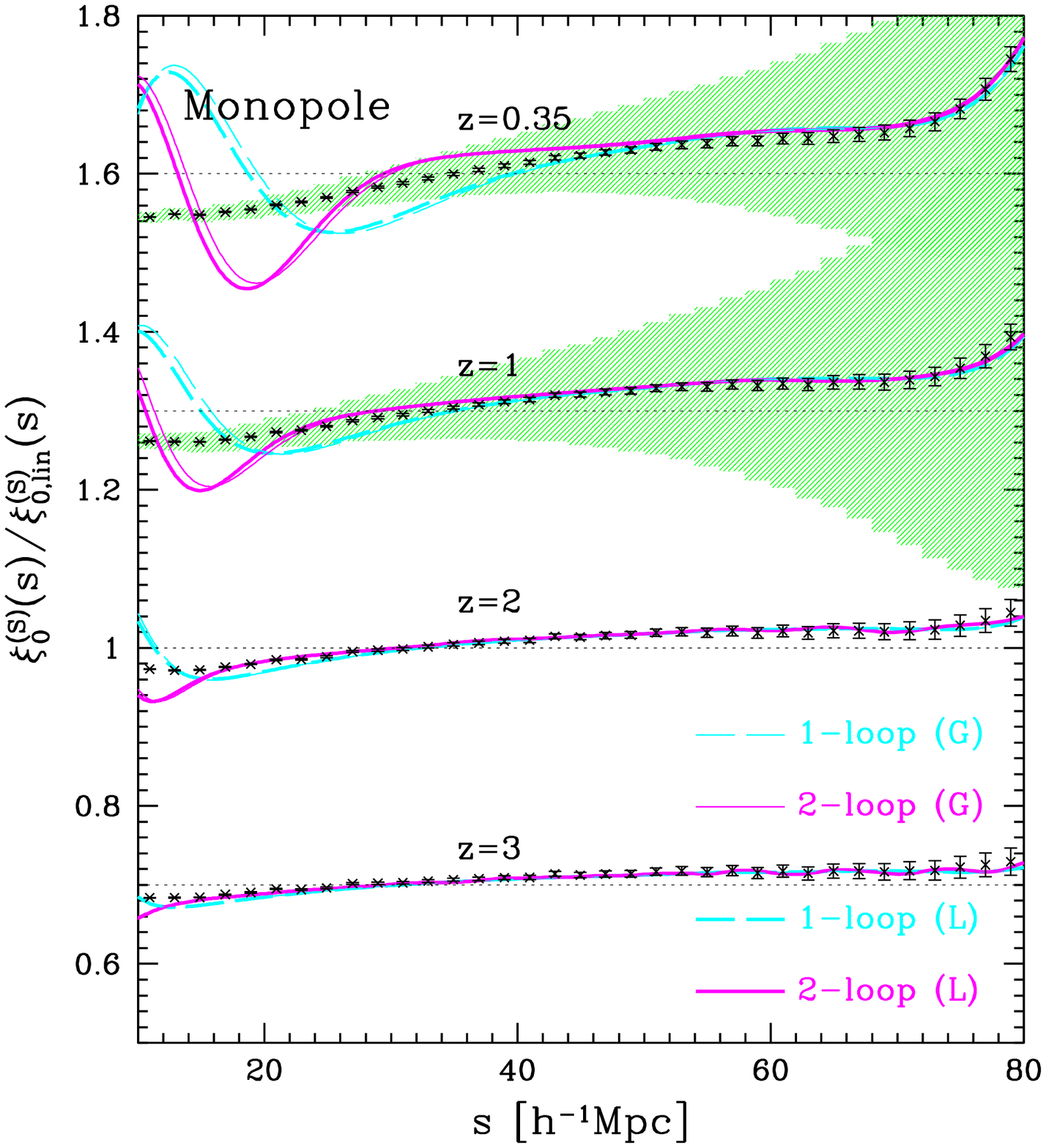}
\hspace*{-0.8cm}
\includegraphics[width=6.6cm,angle=0]{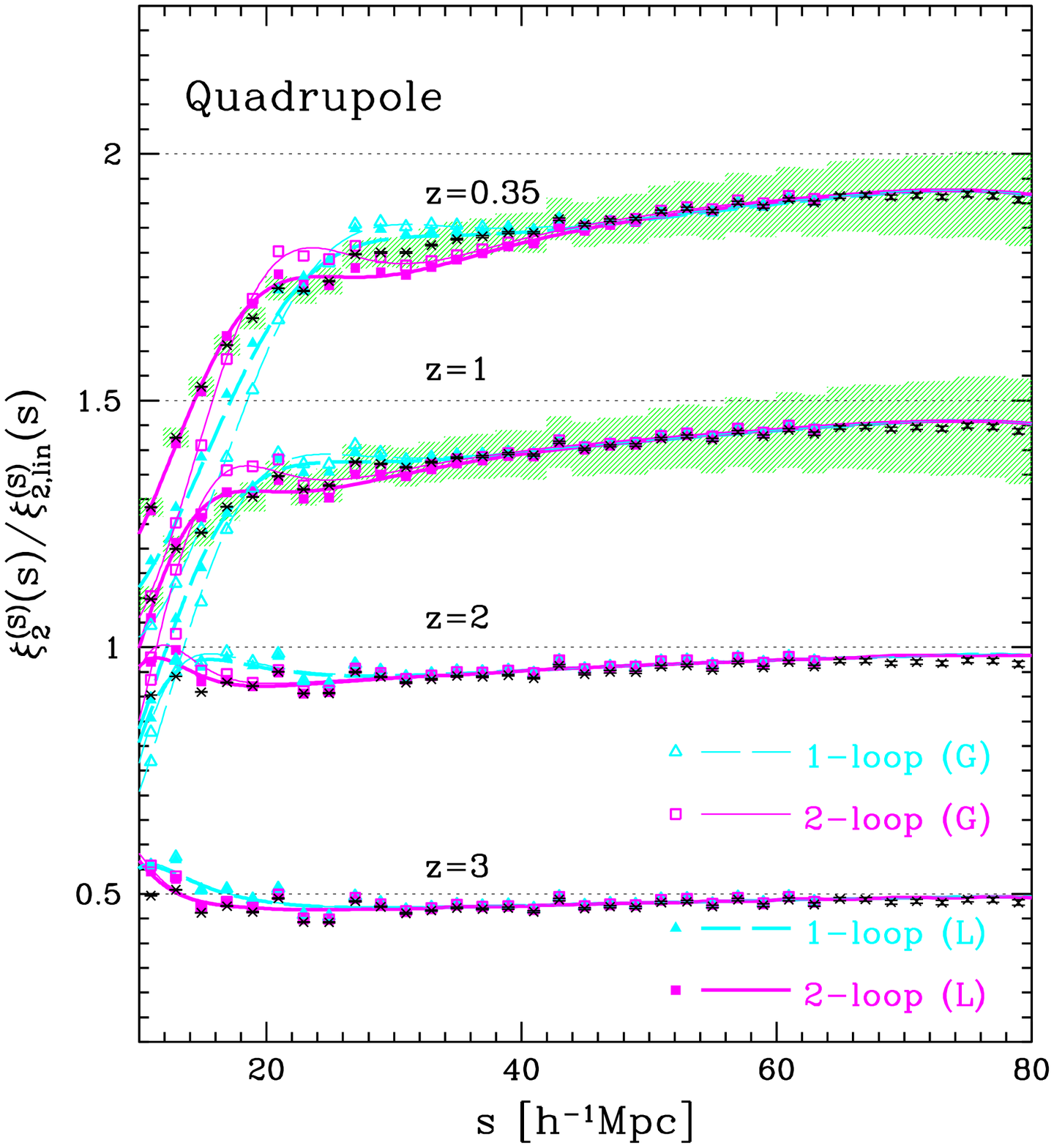}
\hspace*{-0.8cm}
\includegraphics[width=6.6cm,angle=0]{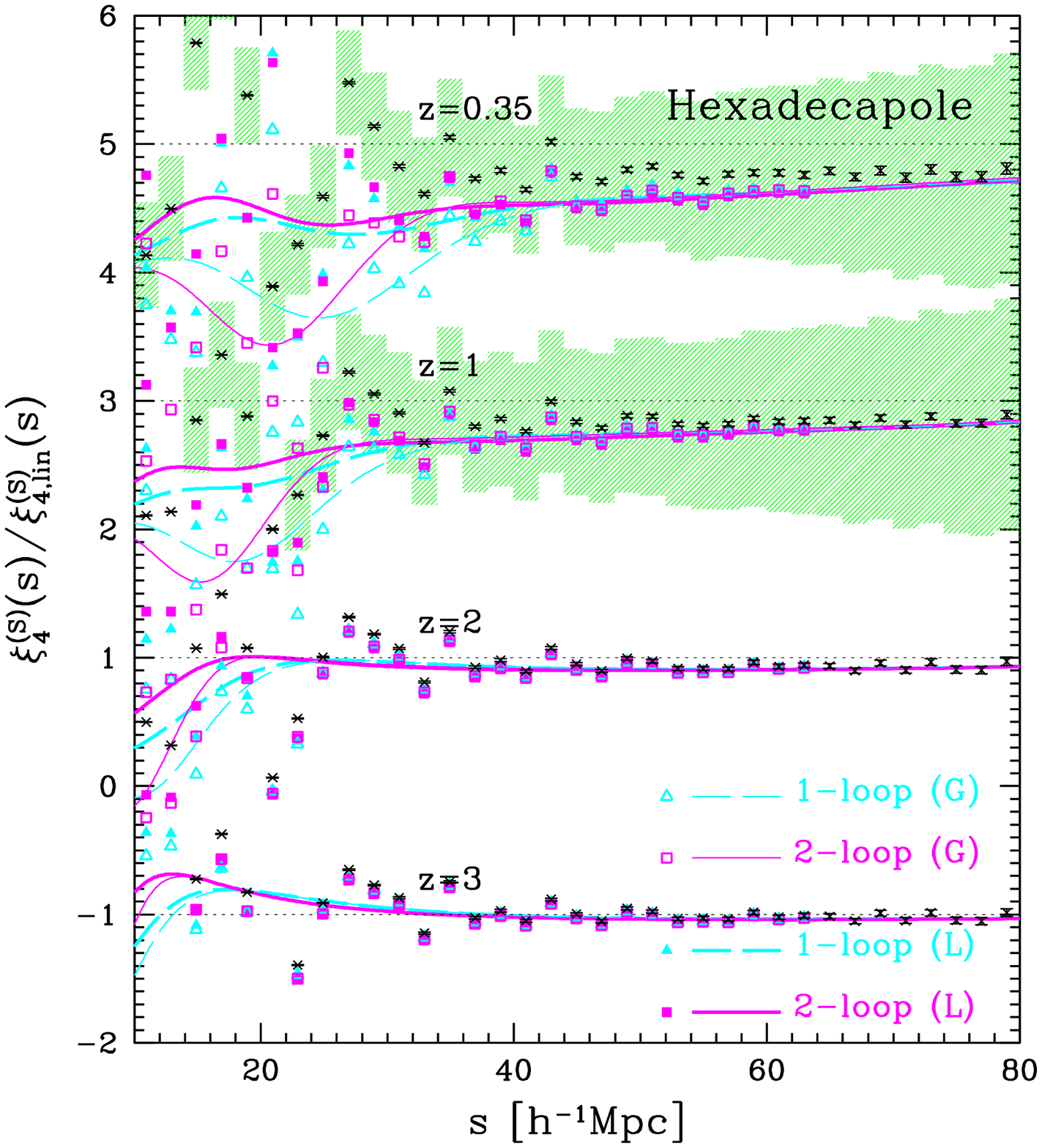}
\end{center}

\vspace*{-0.5cm}

\caption{Redshift-space correlation functions at small scales. Plotted 
  results are the ratio of correlation function to the linear theory 
  predictions taking account of the linear Kaiser factor, i.e., 
  $\xi_\ell^{\rm(S)}(s)/\xi_{\ell,{\rm lin}}^{\rm(S)}(s)$. For clarity, 
  we artificially shift the results at each redshift by a constant value 
  (indicated by the horizontal dotted line). Symbols and line types are 
  the same as those in Fig.~\ref{fig:ratio_pk_red_PT}. The green 
  shaded regions at $z=0.35$ and $1$ indicate 
  the expected $1$-$\sigma$ error of the hypothetical galaxy survey with 
  the volume $V=5\,h^{-3}$\,Gpc$^3$ and number density 
  $n=5\times10^{-4}\,h^3$\,Mpc$^{-3}$. 
\label{fig:xi_red_PT2}}
\end{figure*}

We next consider the correlation function. We first show in 
Fig.~\ref{fig:xi_red_PT1}  the 
the large-scale behavior of the correlation functions, focusing  on scales 
around the baryon acoustic peak. Left, middle and right panels 
respectively plot the results of monopole, quadrupole, and hexadecapole 
correlation functions. The predictions depicted as solid and dashed lines 
are basically obtained from the power spectra through the relation:
\begin{align}
\xi_\ell^{\rm(S)}(s) = i^\ell\int\frac{dk\,k^2}{2\pi^2}\,P_\ell^{\rm(S)}(k)\,
j_\ell(ks).
\end{align}
Note again that thin and thick lines represent the predictions adopting 
the Gaussian and Lorentzian damping functions, and 
we use the same velocity dispersion $\sigmav$ as determined 
in the power spectrum analysis. 
At all redshifts, the one- and two-loop results do indeed agree with 
$N$-body results quite well. 
This is to be contrasted with previous 
studies neglecting $A$ and $B$ terms \cite{Taruya:2009ir}, in which  
the power spectra, $\Pdd$, $\Pdv$ and $\Pvv$, are 
computed with closure theory.  
Now with the coherent treatment with RegPT, the model successfully 
describes the correlation functions around the baryon acoustic peak.  
The results show that at large-scales, the choice of the damping function 
hardly change the prediction, and both the one- and two-loop 
predictions almost coincide each other.

Let us look at the small-scale behaviors 
beyond the baryon acoustic scales. 
Fig.~\ref{fig:xi_red_PT2} shows the ratio of the 
correlation functions to the linear theory predictions, 
$\xi^{\rm(S)}_\ell(s)/\xi^{\rm(S)}_{\ell,{\rm lin}}(s)$, specifically 
focusing on the scales $10\,h^{-1}$\,Mpc\,$\leq s \leq$ $80\,h^{-1}$\,Mpc. 
Note that the linear theory prediction $\xi^{\rm(S)}_{\ell,{\rm lin}}$ 
is made with the linear power spectrum taking 
only account of the linear Kaiser effect. As references, 
we also consider the hypothetical galaxy survey, and 
show the $1$-$\sigma$ statistical errors at 
$z=0.35$ and $1$, depicted as green shaded region. 
This is estimated from 
\begin{align}
[\Delta \xi_\ell^{\rm(S)}(s)]^2=\frac{2}{V} 
\int \frac{dk\,k^2}{2\pi^2}\left\{j_\ell(k\,s)\,\sigma_{P,\ell}(k)\right\}^2
\end{align}
with $\sigma_{P,\ell}$ defined in Eq.~(\ref{eq:sigma_P}). 
Here, we adopt the same survey parameters as we considered in 
Fig.~\ref{fig:ratio_pk_red_PT}.

As anticipated from the power spectrum results, 
the predictions for both the monopole and quadrupole moments 
perfectly agree well with simulations at the scales, 
$s\gtrsim10-30\,h^{-1}$\,Mpc, depending on the redshift. 
The range of agreement with $N$-body 
simulations is comparable to the one obtained in real space, and roughly 
matches the range inferred from the power spectrum results. 
One noticeable point in the prediction of correlation function is 
that even the one-loop results do give an  
accurate prediction over a wide range of correlation function,  
where the choice of damping function hardly change the results.

On the other hand, similar to the power spectrum analysis, 
the measured hexadecapole moment 
suffers from the effect of finite grid-size,  
and in order to make a fair comparison, 
we need to incorporate the effect of this into the theoretical 
calculation.  The triangles and squares are the 
results taking account of the finite grid-size based on the prescription in 
Appendix \ref{App:finite_grid}. Then, the predictions at high-$z$ 
reproduce the $N$-body results almost perfectly, while we find a 
systematic discrepancy at low-$z$, where the results also show a sensitive 
dependence on the choice of the damping function. 
However, we note that the discrepancy seen in the correlation function is 
smaller than the statistical errors of the hypothetical surveys, and it seems 
less significant compared to the results in power spectrum. 
This is partly because many Fourier modes 
can contribute to the correlation function, and they help to 
mitigate the significance of the discrepancy seen in the 
power spectrum at some specific modes. Strictly speaking, 
the amplitudes of the correlation function are strongly correlated between 
different scales, and the error covariance of the correlation function 
may be important to rigorously judge the significance of the discrepancy.  
Rather, a conservative and generic statement is that 
for the scales where both the one- and two-loop 
predictions coincide each other, the non-linear effect of RSD 
and gravity can be small, and thus the discrepancy 
between the PT predictions and simulations is insignificant.

\section{Impact of higher-order corrections}
\label{sec:discussion}

\begin{figure}

\vspace*{-2cm}

\begin{center}
\includegraphics[width=9cm,angle=0]{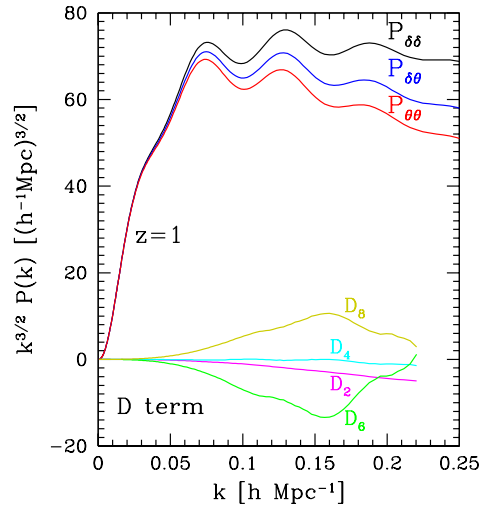}
\end{center}

\vspace*{-0.2cm}

\caption{Same as in Fig.~\ref{fig:pkcorr_A_B}, but 
the scale-dependent coefficient of the $D$ term at $z=1$ is plotted 
($D_2$: magenta, $D_4$: cyan, $D_6$: green, $D_8$: 
yellow). For reference, the power spectra $P_{\delta\delta}$, $P_{\delta\theta}$, 
and $P_{\theta\theta}$ computed from the RegPT at two-loop order 
are also shown. 
\label{fig:pkred_D_term}}
\end{figure}

So far, we have neglected the contribution of the $D$ term 
[Eq.~(\ref{eq:D})] to the predictions of redshift-space power spectrum. 
Strictly speaking, this is no longer valid when we perform the 
two-loop calculations. The higher-order corrections of the RSD 
such as the $D$ term may 
help us to explain the small discrepancy seen in the hexadecapole 
power spectrum. 
In this section, we quantitatively compute the $D$ term, and 
discuss the validity of the analysis in previous section. Further, 
using the measured power spectrum of the $N$-body simulation, 
we also estimate the magnitude of a possible systematics coming from 
the higher-order corrections to the RSD.

Let us first consider the $D$ term. Eq.~(\ref{eq:D}) can be recast as 
\begin{align}
& D(k,\mu)= (k\mu\,f)^2\int\frac{d^3\bfp d^3\bfq}{(2\pi)^6}\,
\frac{p_zq_z}{(pq)^2}
\nonumber\\
&\quad\times
\left\{
T_\sigma(\bfp,\bfq,-\bfk-\bfp-\bfq,\bfk)
-T_\sigma(\bfp,\bfq,\bfk-\bfp,-\bfk-\bfq)
\right\}, 
\label{eq:D_term}
\end{align}
where the function $T_\sigma$ is the cross trispectrum defined by 
\begin{align}
&\left\langle \theta(\bfk_1)\langle \theta(\bfk_2)
\left\{\delta(\bfk_3)+f\,\frac{k_{3z}^2}{k_3^2}\theta(\bfk_3)\right\}
\left\{\delta(\bfk_4)+f\,\frac{k_{4z}^2}{k_4^2}\theta(\bfk_4)\right\}
\right\rangle_c
\nonumber\\
&\quad
=(2\pi)^3\delta_D(\bfk_1+\bfk_2+\bfk_3+\bfk_4)\,
T_\sigma(\bfk_1,\bfk_2,\bfk_3,\bfk_4).
\label{eq:def_T_sigma}
\end{align}
In deriving the above expression, we
have used the symmetric properties of the trispectrum $T_\sigma$, i.e.,  
$T_\sigma(\bfk_1,\bfk_2,\bfk_3,\bfk_4)=
T_\sigma(-\bfk_1,-\bfk_2,-\bfk_3,-\bfk_4)=
T_\sigma(\bfk_2,\bfk_1,\bfk_3,\bfk_4)=
T_\sigma(\bfk_1,\bfk_2,\bfk_4,\bfk_3)$.
The trispectrum $T_\sigma$ is rewritten with the sum of the 
cross power spectrum $T_{abcd}$ for the doublet $\Psi_a$ as
\begin{align}
&T_\sigma(\bfk_1,\bfk_2,\bfk_3,\bfk_4)
=T_{2211}(\bfk_1,\bfk_2,\bfk_3,\bfk_4)
\nonumber\\
&\qquad\qquad\qquad
+f\,\left(\frac{k_{3,z}}{k_3}\right)^2\,T_{2221}(\bfk_1,\bfk_2,\bfk_3,\bfk_4)
\nonumber\\
&\qquad\qquad\qquad
+f\,\left(\frac{k_{4,z}}{k_4}\right)^2\,T_{2212}(\bfk_1,\bfk_2,\bfk_3,\bfk_4)
\nonumber\\
&\qquad\qquad\qquad
+f^2\,\left(\frac{k_{3,z}k_{4,z}}{k_3 k_4}\right)^2\,
T_{2222}(\bfk_1,\bfk_2,\bfk_3,\bfk_4), 
\label{eq:T_sigma_Tabcd}
\end{align}
which can be computed with RegPT. For the predictions of redshift-space 
power spectrum at two-loop order, it is sufficient to give the 
tree-level results for $T_{abcd}$, whose explicit expression is 
given in Appendix \ref{app:tk_RegPT}.

Fig.~\ref{fig:pkred_D_term} shows the 
$D$ term computed at specific redshift $z=1$. 
Similar to the $A$ and $B$ terms, 
the $D$ term can be 
expanded in powers of $\mu$ 
[see Eqs.~(\ref{eq:D_term}) and (\ref{eq:T_sigma_Tabcd})]: 
$D(k,\mu)=\sum_{n=1}^4\,\mu^{2n}\,D_{2n}(k)$.  
We then plot the coefficients $D_{2n}$ as function of wavenumber.  
Note that unlike the 
$A$ and $B$ terms, we were not able to derive a simpler 
expression like Eqs.~(\ref{eq:A_term_simplified}) and (\ref{eq:B_PT_formula}). 
Hence, we employed the Monte Carlo technique to 
directly perform the six-dimensional integral in Eqs.~(\ref{eq:D_term}), 
and obtained the result in $(k,\mu)$-plane. For each wavenumber, we applied
the multipole expansion, and  
characterize the $\mu$-dependence of the $D$ term by the Legendre 
polynomials. Finally, the resultant coefficients are 
translated into those in the power-law expansion, $D_{2n}$.

\begin{figure}

\begin{center}
\includegraphics[width=8cm,angle=0]{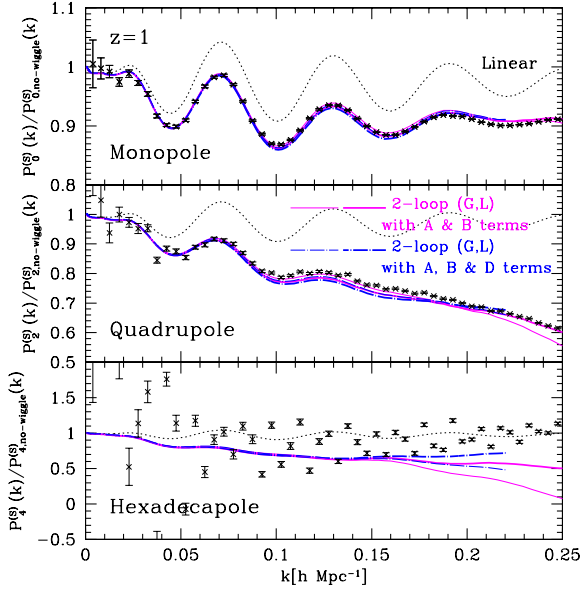}
\end{center}

\vspace*{-0.3cm}

\caption{Ratio of power spectra to the smoothed 
reference spectra in redshift space, 
  $P_\ell^{\rm (S)}(k)/P_{\ell,{\rm no\mbox{-}wiggle}}^{\rm (S)}(k)$ for the predictions
including the $D$ terms (blue dot-dashed). 
The results at $z=1$ are specifically shown.  For comparison, 
the two-loop results ignoring the $D$ terms are also shown 
(magenta solid). Thin and thick lines are the results adopting the 
Gaussian and Lorentzian damping function, respectively. 
\label{fig:ratio_pkred_Dterm}}
\end{figure}

The $D$ term shown in Fig.~\ref{fig:pkred_D_term} 
has the amplitudes roughly comparable to the $B$ term. 
However, a careful look at 
the $\mu$-dependence reveals that the coefficients $D_2$ and $D_4$ are 
rather small. Also, while the amplitude is non-negligible for $D_6$ and 
$D_8$, their signs are opposite each other. 
These facts imply that when convolving with the damping function $D_{\rm FoG}$, 
the contribution of the $D$ term becomes negligible for the monopole and 
quadrupole spectra, and is largely canceled for the hexadecapole power 
spectrum. This is indeed manifested in 
Fig.~\ref{fig:ratio_pkred_Dterm}, 
where we compare the model prediction including the $D$ term (blue dot-dashed) 
with $N$-body simulations. Except for a slight change in the
hexadecapole, which makes the prediction slightly better if we adopt the 
Lorentzian damping function, the resultant power spectra 
are hardly affected by 
the $D$ term. The fitted value of the parameter $\sigmav$ almost remains 
the same. Therefore, we conclude that the actual contribution of the 
$D$ term is less significant for the prediction of 
redshift-space power spectrum.

\begin{figure}
\begin{center}
\hspace*{-0.5cm}
\includegraphics[width=9.2cm,angle=0]{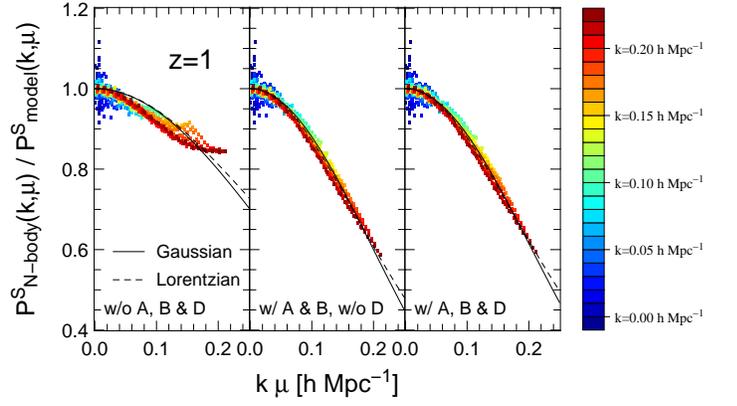}
\end{center}

\vspace*{-0.5cm}

\caption{The ratio of power spectra measured in $(k,\mu)$ space from 
  the simulations to those from the theoretical calculation
[see Eq.(\ref{eq:ratio_pkred})]. The results at $z=1$ 
are plotted as function of $k\mu$. The color scale indicate 
the power spectrum contribution at different range of wavenumbers. Left panel 
shows the results ignoring 
all the corrections, $A$, $B$ and $D$ terms, in the theoretical calculation. 
The middle and right panels are the results taking account of the corrections, 
but in middle panel, the $D$ term is neglected. For reference, we also show 
the Gaussian ($D_{\rm FoG}=\exp\{-(k\,\mu\,f\,\sigmav)^2\}$, solid) and 
Lorentzian ($D_{\rm FoG}=1/\{1+(k\,\mu\,f\,\sigmav)^2/2\}^2$, dashed) 
damping functions adopting the fitted values of $\sigmav$. 
\label{fig:FoG}}
\end{figure}

Nevertheless, this does not prove that the corrections higher 
than the $D$ term originated from the low-$k$ expansion in 
Eq.~(\ref{eq:low-k_expansion}) are entirely negligible. Further, as we saw 
in previous section, there appears a small but non-negligible difference 
in the hexadecapole spectrum between the model predictions and 
$N$-body simulations (Fig.~\ref{fig:ratio_pk_red_PT}). Since  
a part of our RSD model has not been derived by the perturbative expansion,  
there would exist some missing corrections that can systematically affect 
the prediction of higher-multipole power spectra.

In order to elucidate the potential systematics, we measure the 
power spectrum of $N$-body simulations in $(k,\mu)$ 
space, and divide it by the PT prediction in the following way:
\begin{align}
\frac{P_{\rm N\mbox{-}body}^{\rm(S)}(k,\mu)}
{P_{\rm Kaiser}(k,\mu)+A(k,\mu)+B(k,\mu)+D(k,\mu)}
\label{eq:ratio_pkred}
\end{align}
In Fig.~\ref{fig:FoG},  
the results at $z=1$ are plotted as the function of the single 
variable, $k\mu$. The color scales represent the contributions 
from the different wavenumbers. The plotted results are 
the residual contributions of the RSD that is not described by the low-$k$ 
expansion, but is rather characterized by the damping function $D_{\rm FoG}$ 
in our PT model [Eq.~(\ref{eq:TNS_model})]. 
Compared to the case ignoring the correction 
terms (left panel), the residuals shown in the middle and right panels are 
mostly characterized by the single-valued function of $k\mu$, and 
within the plotted range, 
they are approximately described by the Gaussian 
($D_{\rm FoG}=e^{-(k\mu f \sigmav)^2}$, solid lines) or Lorentzian 
($D_{\rm FoG}=1/\{1+(k\mu f\sigmav)^2/2\}^2$, dashed lines) damping function. 
This is indeed what we expected from our RSD model.

However, a closer look at 
the scatter around the damping functions 
reveals some systematics that  
the low-$k$ residuals tend to take larger values, 
while the high-$k$ residuals 
take slightly smaller values than the Gaussian or Lorentzian damping function, 
indicating the imperfect modeling of RSD. 
These systematics in the scatter would be certainly the 
source of the discrepancy seen in the hexadecapole power spectrum. 
In this respect, a better description for the power spectrum suppression
might be crucial 
for an accurate prediction of higher-multipole spectra. Recent study
by Ref.~\cite{Vlah:2012ni} suggests that a simple prescription with the 
function $D_{\rm FoG}$ might be a oversimplified treatment, 
and the suppression of power spectrum cannot be controlled by a single 
parameter $\sigmav$. These two facts may pose a caution for the 
assumption of our model, and the suppression and modulation of the power 
spectrum might not be treated separately. Thus, in general, 
a more elaborate modeling of RSD is needed for an accurate prediction of 
higher multipole power spectra, although the actual impact 
of such a modeling on the cosmological application is unclear,  
and is beyond the scope of this paper.

\section{Conclusion}
\label{sec:conclusion}

In this paper, we have implemented the resummed PT scheme called RegPT,
which based on a multi-point propagator expansion,
together with an improved model of  redshift-space distortions (RSD), in order to compute
the matter two-point statistical properties in redshift space, both in Fourier and in configuration 
spaces. For the first time, calculations consistently include 
PT corrections up to the two-loop order. 
Our analytical predictions are then  compared with $N$-body simulations. We found 
that, whenever predictions are at percent level in real space, a similar performance can be obtained in redshift space. With a full implementation of 
the RegPT scheme, the present work puts forward proper PT calculations 
in redshift space, allowing the predictions to be greatly 
improved in the sense that we are now able to 
give a consistent prediction not only for the power spectrum but also for 
the correlation function. This is in marked contrast to  
the previous analysis partly using the standard PT treatment.

We have also investigated the influence of higher-order 
corrections in redshift space. While our RSD model has been derived 
based on the 
low-$k$ expansion, 
the previous studies have missed the non-Gaussian correction at the trispectrum 
order called $D$ term, whose contribution is comparable to the two-loop 
order. We have computed the $D$ term explicitly, and confirmed that the 
amplitude of the coefficients given in powers of $\mu$ 
are comparable to that of the $B$ term. 
The actual contribution to the power spectrum, however, turns out to be 
small and it would only affect the higher-multipole spectra. Nevertheless, 
the present RSD model seems to have a slight deficit in the prediction
of the higher multipoles, and we have quantified the 
validity of our model assumption with a help of $N$-body 
simulations.  
The results in Fig.~\ref{fig:FoG} indicate an improper treatment for the 
damping effect of the power spectrum, and this would lead to the 
small discrepancy of the higher multipoles between predictions  
and simulations.

Nevertheless, the actual impact of the improper modeling of RSD might 
be less significant, considering the other big systematics such as  
the galaxy bias. Indeed, one crucial aspect of the galaxy bias in 
redshift space 
is that it can affect both the density and velocity fields, and 
the clustering feature in redshift space cannot be straightforwardly 
understood from the real-space clustering. We have previously shown
that our model of RSD, combined with the simple linear bias,  
can successfully describe the redshift-space halo clustering in $N$-body 
simulations quite well. Still, however, this should be regarded as 
an idealistic example. In reality, galaxies do not faithfully trace the 
halo clustering, and peculiar velocity of galaxies certainly differs 
from the center-of-mass velocity of halos. As it has been recently  
advocated in Ref.~\cite{Hikage:2011ut}, off-centered galaxies can have 
a large peculiar velocity due to the virial motion, and the presence of 
these galaxies can drastically change the power spectrum 
\cite{Hikage:2012zk,Masaki:2012gh}. The significance 
of this effect may sensitively depend on the type of the targeted galaxies. 
A careful study of the galaxy samples is quite crucial for the 
cosmological analysis to get an unbiased cosmological constraint.

\begin{acknowledgments}
This work has been benefited from exchange visits supported by a 
bilateral grant from 
Minist\`ere  Affaires Etrang\`eres et Europ\'eennes in France
and Japan Society for the Promotion of Science (JSPS).  
A.T. is supported in part by a Grant-in-Aid for Scientific 
Research from the JSPS (No.~24540257). 
T. N. is supported by a Grant-in-Aid for JSPS Fellows (PD: 22-181) and 
by World Premier International Research Center Initiative
(WPI Initiative), MEXT, Japan. Numerical computations for the present work 
have been
carried out in part on Cray XT4 at Center for Computational Astrophysics, 
CfCA, of National
Astronomical Observatory of Japan, and in part under the Interdisciplinary 
Computational 
Science Program in Center for Computational Sciences, University of Tsukuba.
F.B is also partly supported by the French Programme National 
de Cosmologie et Galaxies. 
\end{acknowledgments}

\appendix
\section{Explicit expressions for regularized power spectrum, bispectrum and trispectrum}
\label{app:Gamma_expansion_Pk_Bk}

In this Appenix, based on the regularized $\Gamma$ expansion, 
we give full expressions for the real-space 
power spectra and bispectra used to compute the 
redshift-space power spectrum and correlation function.

\subsection{Power spectrum}
\label{subsec:pk_RegPT}

According to the prescription described in 
Sec.~\ref{sec: pk_bk_from_Gamma_expansion}, 
the power spectrum at one-loop order of the $\Gamma$-expansion 
becomes
\begin{widetext}
\begin{align}
&P_{ab}(k;\eta) = \Gamma_{a,{\rm reg}}^{(1)}(k;\eta)
\Gamma_{b,{\rm reg}}^{(1)}(k;\eta)P_0(k)+
2\int\frac{d^3\bfq}{(2\pi)^3}\,
\Gamma_{a,{\rm reg}}^{(2)}(\bfq,\bfk-\bfq;\eta)
\Gamma_{b,{\rm reg}}^{(2)}(\bfq,\bfk-\bfq;\eta)P_0(q)P_0(|\bfk-\bfq|)
\label{eq:pk_Gamma_reg_1loop}
\end{align}
with the regularized propagators $\Gamma^{(1)}_{\rm reg}$ and 
$\Gamma^{(2)}_{\rm reg}$ valid at one-loop order being: 
\begin{align}
&\Gamma_{a,{\rm reg}}^{(1)}(k;\eta)=e^{\eta}\left[
1+\frac{k^2\sigmav^2e^{2\eta}}{2}+
e^{2\eta}\,\overline{\Gamma}^{(1)}_{a,{\rm 1\mbox{-}loop}}(k)\right]
\exp\left\{-\frac{k^2\sigmav^2e^{2\eta}}{2}\right\}, 
\label{eq:Gamma1_reg_1loop}
\\
& \Gamma_{a,{\rm reg}}^{(2)}(\bfq,\bfk-\bfq;\eta)=e^{2\eta}
F_{a}^{(2)}(\bfq,\bfk-\bfq)\,
\exp\left\{-\frac{k^2\sigmav^2e^{2\eta}}{2}\right\}.
\label{eq:Gamma2_reg_tree}
\end{align}
\end{widetext}

On the other hand, if we include the next-to-leading order contribution, 
i.e., two-loop corrections, the resultant expression becomes
\begin{widetext}
\begin{align}
&P_{ab}(k;\eta) = 
\Gamma_{a,{\rm reg}}^{(1)}(k;\eta)\Gamma_{b,{\rm reg}}^{(1)}(k;\eta)P_0(k)+
2\int\frac{d^3\bfq}{(2\pi)^3}\,
\Gamma_{a,{\rm reg}}^{(2)}(\bfq,\bfk-\bfq;\eta)
\Gamma_{b,{\rm reg}}^{(2)}(\bfq,\bfk-\bfq;\eta)P_0(q)P_0(|\bfk-\bfq|)
\nonumber\\
&\qquad\qquad\qquad\qquad+6\int\frac{d^6\bfp d^3\bfq}{(2\pi)^6}\,
\Gamma_{a,{\rm reg}}^{(3)}(\bfp,\bfq,\bfk-\bfp-\bfq;\eta)
\Gamma_{b,{\rm reg}}^{(3)}(\bfp,\bfq,\bfk-\bfp-\bfq;\eta)
P_0(p)P_0(q)P_0(|\bfk-\bfp-\bfq|)
\label{eq:pk_Gamma_reg_2loop}
\end{align}
with the regularized propagators given by
\begin{align}
&\Gamma_{a,{\rm reg}}^{(1)}(k;\eta)=e^{\eta}\left[
1+
\frac{k^2\sigmav^2e^{2\eta}}{2}+
\frac{1}{2}\left(\frac{k^2\sigmav^2e^{2\eta}}{2}\right)^2
+e^{2\eta}\,\overline{\Gamma}^{(1)}_{a,{\rm 1\mbox{-}loop}}(k)
\left\{1+\frac{k^2\sigmav^2e^{2\eta}}{2}\right\}+
e^{4\eta}\,\overline{\Gamma}^{(1)}_{a,{\rm 2\mbox{-}loop}}(k)\right]
\exp\left\{-\frac{k^2\sigmav^2e^{2\eta}}{2}\right\},
\label{eq:Gamma1_reg}
\\
& \Gamma_{a,{\rm reg}}^{(2)}(\bfq,\bfk-\bfq;\eta)=e^{2\eta}\left[
F_{a}^{(2)}(\bfq,\bfk-\bfq)\left\{1+
\frac{k^2\sigmav^2e^{2\eta}}{2}\right\}+
e^{2\eta}\,\overline{\Gamma}^{(2)}_{a,{\rm 1\mbox{-}loop}}(\bfq,
\bfk-\bfq)\right]
\exp\left\{-\frac{k^2\sigmav^2e^{2\eta}}{2}\right\},
\label{eq:Gamma2_reg}
\\
&\Gamma_{a,{\rm reg}}^{(3)}(\bfp,\bfq,\bfk-\bfp-\bfq;\eta)=
e^{3\eta}\,F_{a}^{(3)}(\bfp,\bfq,\bfk-\bfp-\bfq)
\exp\left\{-\frac{k^2\sigmav^2e^{2\eta}}{2}\right\}, 
\label{eq:Gamma3_reg}
\end{align}
\end{widetext}
where the quantity $\overline{\Gamma}_{p\mbox{-}{\rm loop}}^{(n)}$ is defined in 
Eq.~(\ref{eq:Gamma-p_nloop}). 
The higher-order contributions up to the two- and one-loop order
of the propagators are respectively included in the expression 
of the regularized propagators $\Gamma_{a,{\rm reg}}^{(1)}$ and 
$\Gamma_{a,{\rm reg}}^{(2)}$,  consistently with the 
$\Gamma$-expansion at two-loop order. The detailed prescription on how to 
efficiently compute each contribution is described in 
Ref.~\cite{Taruya:2012ut}. Shortly, we use the analytic expression of 
the kernels to compute $\overline{\Gamma}_{p\mbox{-}{\rm loop}}^{(n)}$ 
($\overline{\Gamma}_{1\mbox{-}{\rm loop}}^{(2)}$ for 
\cite{Bernardeau:2011dp} and $\overline{\Gamma}_{2\mbox{-}{\rm loop}}^{(1)}$ 
for \cite{Bernardeau:2012ux}), and 
the integrals are performed with the method of Gaussian quadrature. For
the integral in the last term of Eq.~(\ref{eq:pk_Gamma_reg_1loop}), 
we adopt the Monte Carlo technique using the publicly available library 
called CUBA \cite{Hahn:2004fe}.

\subsection{Bispectrum}
\label{subsec:bk_RegPT}

The real-space bispectra are needed to compute the $A$ term, and for a 
consistent calculation of the redshift-space power spectrum up to two-loop 
order, we need the bispectra at one-loop order. In this case, we can 
decomposed the bispectrum into three pieces: 
\begin{align}
&  B_{abc}(\bfk_1,\bfk_2,\bfk_3) = B_{abc}^{I}(\bfk_1, \bfk_2, \bfk_3) 
\nonumber\\
&\qquad
+B_{abc}^{II}(\bfk_1, \bfk_2, \bfk_3) + B_{abc}^{III}(\bfk_1, \bfk_2, \bfk_3).   
\label{eq:bispectrum_I_II_III}
\end{align}
These three contributions are 
expressed in terms of the multi-point propagator $\Gamma^{(p)}$ 
\cite{Bernardeau:2011dp}: 
\begin{widetext}
\begin{align}
&  B_{abc}^{I}(\bfk_1,\bfk_2,\bfk_3) = 2\,\left\{
\Gamma^{(2)}_{a,{\rm reg}}(\bfk_2,\bfk_3)
\Gamma_{b,{\rm reg}}^{(1)}(k_2)\Gamma_{c,{\rm reg}}^{(1)}(k_3)\,P_0(k_2)P_0(k_3)
\right.
\nonumber\\
&\left.\qquad\qquad+
\Gamma^{(2)}_{b,{\rm reg}}(\bfk_1,\bfk_3)\Gamma_{a,{\rm reg}}^{(1)}(k_1)
\Gamma_{c,{\rm reg}}^{(1)}(k_3)\,P_0(k_1)P_0(k_3)+
\Gamma^{(2)}_{c,{\rm reg}}(\bfk_1,\bfk_2)\Gamma_{a,{\rm reg}}^{(1)}(k_1)
\Gamma_{b,{\rm reg}}^{(1)}(k_2)\,P_0(k_1)P_0(k_2)
\right\},
\\
&  B_{abc}^{II}(\bfk_1,\bfk_2,\bfk_3) = 8\,\int\frac{d^3\bfq}{(2\pi)^3}\,
\Gamma_{a,{\rm reg}}^{(2)}(\bfk_1-\bfq,\bfq)\,
\Gamma_{b,{\rm reg}}^{(2)}(\bfk_2+\bfq,-\bfq)\,
\Gamma_{c,{\rm reg}}^{(2)}(-\bfq-\bfk_2,-\bfk_1+\bfq)\,
\nonumber\\
&\qquad\qquad\qquad\qquad\qquad\qquad\times\,
P_0(|\bfk_1-\bfq|)P_0(|\bfk_2+\bfq|)P_0(q),
\\
&  B_{abc}^{III}(\bfk_1,\bfk_2,\bfk_3) = 6\,\int\frac{d^3\bfq}{(2\pi)^3}\,
\left\{ 
\Gamma_{a,{\rm reg}}^{(3)}(-\bfk_3,-\bfk_2+\bfq,-\bfq)\,
\Gamma_{b,{\rm reg}}^{(2)}(\bfk_2-\bfq,\bfq)\,
\Gamma_{c,{\rm reg}}^{(1)}(k_3)\,\,P_0(|\bfk_2-\bfq|)P_0(q)P_0(k_3)\right.
\nonumber\\
&
\qquad\qquad\qquad\qquad\qquad\qquad~
+\Gamma_{a,{\rm reg}}^{(3)}(-\bfk_2,-\bfk_3+\bfq,-\bfq)\,
\Gamma_{b,{\rm reg}}^{(1)}(k_2)\,
\Gamma_{c,{\rm reg}}^{(2)}(\bfk_3-\bfq,\bfq)\,\,P_0(|\bfk_3-\bfq|)P_0(q)P_0(k_2)
\nonumber\\
&
\qquad\qquad\qquad\qquad\qquad\qquad~
+\Gamma_{a,{\rm reg}}^{(2)}(\bfk_1-\bfq,\bfq)\,
\Gamma_{b,{\rm reg}}^{(3)}(-\bfq,-\bfk_1+\bfq,-\bfk_3)\,
\Gamma_{c,{\rm reg}}^{(1)}(k_3)\,\,P_0(|\bfk_1-\bfq|)P_0(q)P_0(k_3)
\nonumber\\
&
\qquad\qquad\qquad\qquad\qquad\qquad~
+\Gamma_{a,{\rm reg}}^{(1)}(k_1)\,
\Gamma_{b,{\rm reg}}^{(3)}(-\bfk_1,-\bfk_3+\bfq,-\bfq)\,
\Gamma_{c,{\rm reg}}^{(2)}(\bfk_3-\bfq,\bfq)\,\,P_0(k_1)P_0(q)P_0(|\bfk_3-\bfq|)
\nonumber\\
&
\qquad\qquad\qquad\qquad\qquad\qquad~
+\Gamma_{a,{\rm reg}}^{(2)}(\bfk_1-\bfq,\bfq)\,
\Gamma_{b,{\rm reg}}^{(1)}(k_2)\,
\Gamma_{c,{\rm reg}}^{(3)}(-\bfk_1+\bfq,-\bfq,-\bfk_2)\,\,
P_0(|\bfk_1-\bfq|)P_0(q)P_0(k_2)
\nonumber\\
&
\qquad\qquad\qquad\qquad\qquad\qquad~\left.
+\Gamma_{a,{\rm reg}}^{(1)}(k_1)\,\Gamma_{b,{\rm reg}}^{(2)}(\bfk_2-\bfq,\bfq)\,
\Gamma_{c,{\rm reg}}^{(3)}(-\bfk_1,-\bfq,-\bfk_2+\bfq)\,\,
P_0(k_1)P_0(q)P_0(|\bfk_2-\bfq|).
\,\right\}
\end{align}
\end{widetext}
For the bispectrum $B_{abc}^I$, 
the regularized multipoint propagators  
valid at one-loop order are computed with Eq.~(\ref{eq:Gamma1_reg_1loop}) 
for $\Gamma^{(1)}_{a,{\rm reg}}$ and 
Eq.~(\ref{eq:Gamma2_reg}) for $\Gamma^{(2)}_{a,{\rm reg}}$.  
On the other hand, for the contributions 
$B_{abc}^{II}$ and $B_{abc}^{III}$, 
the tree-level propagator 
with exponential cutoff is sufficient for a 
consistent calculation at one-loop order. Explicitly, it is given by  
\begin{align}
\Gamma^{(n)}_{a,{\rm reg}}(\bfk_1,\cdots,\,\bfk_n)=
F_a^{(n)}(\bfk_1,\cdots,\,\bfk_n)\,
\exp\left\{-\frac{k^2\sigma_d^2\,e^{2\,\eta}}{2}\right\}
\label{eq:G_reg_tree}
\end{align}
with $k=|\bfk_1+\cdots+\bfk_n|$.

Finally, for the redshift-space power spectrum at one-loop order, 
the relevant contribution in the $A$ term 
is the tree-level results of the bispectra.  
The tree-level bispectrum is computed from 
the first term in Eq.~(\ref{eq:bispectrum_I_II_III}), with 
the regularized propagators given by Eq.~(\ref{eq:G_reg_tree}).

\subsection{Trispectrum}
\label{app:tk_RegPT}

Since the lowest-order contribution to the trispectrum is already 
comparable to the two-loop corrections for the redshift-space power 
spectrum, it is sufficient for our case 
to derive the tree-level expression for the trispectrum. 
The lowest-order expression for the trispectrum $T_{abcd}$ becomes 
\begin{align}
&T_{abcd}(\bfk_1,\bfk_2,\bfk_3,\bfk_4)=
T_{abcd}^{I}(\bfk_1,\bfk_2,\bfk_3,\bfk_4)
\nonumber\\
&\qquad\qquad\qquad\qquad
+T_{abcd}^{II}(\bfk_1,\bfk_2,\bfk_3,\bfk_4)
\end{align}
with the contributions $T_{abcd}^{I}$ and 
$T_{abcd}^{II}$ respectively given by
\begin{widetext}
\begin{align}
&T_{abcd}^{I}(\bfk_1,\bfk_2,\bfk_3,\bfk_4)
\nonumber\\
&=4\,\Bigl[
\Gamma_{a,{\rm reg}}^{(2)}(\bfk_{13},-\bfk_3)\Gamma_{b,{\rm reg}}^{(2)}(-\bfk_{13},-\bfk_4)
P_0(k_{13})P_0(k_3)P_0(k_4)+
\Gamma_{a,{\rm reg}}^{(2)}(\bfk_{14},-\bfk_4)\Gamma_{b,{\rm reg}}^{(2)}(-\bfk_{14},-\bfk_3)
P_0(k_{14})P_0(k_{3})P_0(k_4)
\nonumber\\
&~~
+\Gamma_{a,{\rm reg}}^{(2)}(\bfk_{12},-\bfk_2)\Gamma_{c,{\rm reg}}^{(2)}(-\bfk_{12},-\bfk_4)
P_0(k_{12})P_0(k_2)P_0(k_4)+
\Gamma_{a,{\rm reg}}^{(2)}(\bfk_{14},-\bfk_4)\Gamma_{c,{\rm reg}}^{(2)}(-\bfk_{14},-\bfk_2)
P_0(k_{14})P_0(k_{2})P_0(k_4)
\nonumber\\
&~~
+\Gamma_{a,{\rm reg}}^{(2)}(\bfk_{13},-\bfk_3)\Gamma_{d,{\rm reg}}^{(2)}(-\bfk_{13},-\bfk_2)
P_0(k_{13})P_0(k_3)P_0(k_2)+
\Gamma_{a,{\rm reg}}^{(2)}(\bfk_{12},-\bfk_2)\Gamma_{d,{\rm reg}}^{(2)}(-\bfk_{12},-\bfk_3)
P_0(k_{12})P_0(k_{3})P_0(k_2)
\nonumber\\
&~~
+\Gamma_{c,{\rm reg}}^{(2)}(\bfk_{13},-\bfk_1)\Gamma_{b,{\rm reg}}^{(2)}(-\bfk_{13},-\bfk_4)
P_0(k_{13})P_0(k_1)P_0(k_4)+
\Gamma_{c,{\rm reg}}^{(2)}(\bfk_{34},-\bfk_4)\Gamma_{b,{\rm reg}}^{(2)}(-\bfk_{34},-\bfk_1)
P_0(k_{34})P_0(k_{1})P_0(k_4)
\nonumber\\
&~~
+\Gamma_{d,{\rm reg}}^{(2)}(\bfk_{34},-\bfk_3)\Gamma_{b,{\rm reg}}^{(2)}(-\bfk_{34},-\bfk_1)
P_0(k_{34})P_0(k_3)P_0(k_1)+
\Gamma_{d,{\rm reg}}^{(2)}(\bfk_{14},-\bfk_1)\Gamma_{b,{\rm reg}}^{(2)}(-\bfk_{14},-\bfk_3)
P_0(k_{14})P_0(k_{3})P_0(k_1)
\nonumber\\
&~~
+\Gamma_{c,{\rm reg}}^{(2)}(\bfk_{13},-\bfk_1)\Gamma_{d,{\rm reg}}^{(2)}(-\bfk_{13},-\bfk_2)
P_0(k_{13})P_0(k_1)P_0(k_2)+
\Gamma_{c,{\rm reg}}^{(2)}(\bfk_{23},-\bfk_2)\Gamma_{d,{\rm reg}}^{(2)}(-\bfk_{23},-\bfk_1)
P_0(k_{23})P_0(k_{1})P_0(k_2)
\Bigr]
\end{align}
and
\begin{align}
&T_{abcd}^{II}(\bfk_1,\bfk_2,\bfk_3,\bfk_4)=
6\,\Bigl[\Gamma_{a,{\rm reg}}^{(3)}(\bfk_2,\bfk_3,\bfk_4)P_0(k_2)P_0(k_3)P_0(k_4)
+\Gamma_{b,{\rm reg}}^{(3)}(\bfk_1,\bfk_3,\bfk_4)P_0(k_1)P_0(k_3)P_0(k_4)
\nonumber\\
&\qquad\qquad\qquad\qquad\qquad
+\Gamma_{c,{\rm reg}}^{(3)}(\bfk_1,\bfk_2,\bfk_4)P_0(k_1)P_0(k_2)P_0(k_4)
+\Gamma_{d,{\rm reg}}^{(3)}(\bfk_1,\bfk_2,\bfk_3)P_0(k_1)P_0(k_2)P_0(k_3)
\Bigr]
\end{align}
\end{widetext}
with the tree-level multipoint propagator $\Gamma^{(n)}_{a,{\rm reg}}$ 
given in Eq.~(\ref{eq:G_reg_tree}). Note that the expressions for the 
trispectrum given above reduces to those of the standard PT at 
tree order if we just neglect the exponential cutoff in the 
multi-point propagators (e.g., \cite{Scoccimarro:1999kp}).

\section{Effect of finite grid-size in measuring the power spectrum 
and correlation function }
\label{App:finite_grid}

Here, we present the prescription on how to incorporate the effect of 
the finite grid-size into the theoretical prediction of power spectrum 
and correlation function. 

In $N$-body simulations, the measurement of 
the multipole power spectra is done with the density fields assigned on 
grids 
in Fourier space. On each grid, we first evaluate the square of density 
field multiplied by the Legendre polynomial, and take an average over the 
grids within the thin spherical shell around $k$, the width of which 
is given by $\Delta k$. Thus, for a fixed $\Delta k$, 
the number of grids falling 
into the spherical shell inevitably decreases with decreasing wavenumber,  
finally leading to an inhomogeneous sampling. To 
mimic this effect in the theoretical predictions, 
we prepare the same grid space as done in the analysis of the $N$-body data,   
and assign the theoretical power spectrum, given as function of $k_\parallel$ 
and $k_\parallel$, on these grids. Then, multiplying Legendre polynomials, 
we take an average over the spherical thin shell. This is expressed as
\begin{align}
P_\ell^{\rm(S)}(k_i)= \frac{2\ell+1}{N_k} 
\sum_{|\bfk|\in[k_i-\Delta k/2, k_i+\Delta k/2]} 
P^{\rm(S)}(k_\parallel,k_\perp)\,\mathcal{P}_\ell(k_\parallel/k),  
\end{align}
where the quantity $N_k$ is the number of grids falling into the 
spherical thin shell. In the present paper, 
the grid size is chosen as $2\pi/L_{\rm box}$ with box size
$L_{\rm box}=2,048\,h^{-1}$\,Mpc, and the width of the Fourier bin in 
measuring power spectra are set to $\Delta k=0.005\,h$\,Mpc$^{-1}$. 
Note that while the average over the spherical thin shell 
is taken over the domain with positive wavenumber, $k\ge0$, 
we must take care when we sum up the 
contribution from the grids on the boundary of the quadrant. 
To avoid the double counting, the weight factor should be 
appropriately multiplied. For instance, the factor $1/2$ ($1/4$) 
is multiplied when one (two) of the components in wave vector vanishes.

In similar way, the effect of finite grid-size for 
the correlation function can be incorporated into the 
theoretical prediction. Note that this effect 
is prominent only when we adopt the grid-based calculation of  
the correlation function. We express the multipole 
moments of the correlation function 
with the discrete sum over the grids on the configuration space: 
\begin{align}
\xi_\ell^{\rm(S)}(s_i)= \frac{2\ell+1}{N_s} 
\sum_{|\bfs|\in[s_i-\Delta s/2, s_i+\Delta s/2]} 
\xi^{\rm(S)}(s_\parallel,s_\perp)\,\mathcal{P}_\ell(s_\parallel/s),  
\end{align}
where the redshift-space correlation function $\xi^{\rm(S)}$ is 
calculated with 
\begin{align}
\xi^{\rm(S)}(s_\parallel,s_\perp)
&=\int \frac{d^3\bfk}{(2\pi)^3}\,\, 
e^{i\,\bfk\cdot\bfs}\,P^{\rm(S)}(k_\parallel,k_\perp)
\nonumber\\
&=\int \frac{dk_\parallel}{2\pi^2}
\int dk_\perp k_\perp \,P^{\rm(S)}(k_\parallel,k_\perp)
\nonumber\\
&\qquad\qquad\qquad\quad
\times
\cos(k_\parallel s_\parallel)J_0(k_\perp s_\perp). 
\end{align}
For the results shown in Fig.~\ref{fig:xi_red_PT2}, both the grid size 
and the width of bins $\Delta s$ are set to $2\,h^{-1}$\,Mpc 
in configuration space.

\bibliographystyle{apsrev}

\end{document}